\documentclass[preprint,5p,number]{elsarticle}
\usepackage[utf8]{inputenc}
\usepackage{amsmath,commath,amssymb,mathtools}
\usepackage{bm}
\usepackage[hidelinks]{hyperref}
\usepackage[capitalize]{cleveref}
\usepackage{physics}
\usepackage[ruled,vlined]{algorithm2e}
\usepackage{algpseudocode}
\usepackage{csvsimple}
\usepackage[usenames,dvipsnames]{xcolor}
\usepackage{url}
\usepackage{scalerel,stackengine}
\usepackage{siunitx,booktabs}
\usepackage{microtype}  % To prevent several overfull hboxes

\newcommand{\I}{\text{I}}
\newcommand{\II}{\text{II}}

\stackMath
\newcommand\reallywidehat[1]{%
\savestack{\tmpbox}{\stretchto{%
  \scaleto{%
    \scalerel*[\widthof{\ensuremath{#1}}]{\kern.1pt\mathchar"0362\kern.1pt}%
    {\rule{0ex}{\textheight}}%WIDTH-LIMITED CIRCUMFLEX
  }{\textheight}% 
}{2.4ex}}%
\stackon[-6.9pt]{#1}{\tmpbox}%
}

\newcommand{\appropto}{\mathrel{\vcenter{
			\offinterlineskip\halign{\hfil$##$\cr
				\propto\cr\noalign{\kern2pt}\sim\cr\noalign{\kern-2pt}}}}}

\crefname{algocf}{alg.}{algs.}
\Crefname{algocf}{Algorithm}{Algorithms}

\begin{document}
	\title{Vector potential-based MHD solver for non-periodic flows using Fourier continuation expansions}
	
%	\author[1]{Mauro Fontana\corref{cor1}}\ead{mfontana@df.uba.ar}
	\author[1]{Mauro Fontana}\ead{mfontana@df.uba.ar}
	\author[1]{Pablo D. Mininni}
	\author[2]{Oscar P. Bruno}
	\author[1]{Pablo Dmitruk}
%	\cortext[cor1]{Corresponding author.}
	\address[1]{Universidad de Buenos Aires, Facultad de Ciencias Exactas y Naturales, Departamento de Física, \& IFIBA, CONICET, Ciudad Universitaria, Buenos Aires 1428, Argentina.}
	\address[2]{Computing and Mathematical Sciences, Caltech, Pasadena, CA 91125, USA.}
\begin{abstract}
A high-order method to evolve in time electromagnetic and velocity fields in conducting fluids with non-periodic boundaries is presented. The method has a small overhead compared with fast FFT-based pseudospectral methods in periodic domains. It uses the magnetic vector potential formulation for accurately enforcing the null divergence of the magnetic field, and allowing for different boundary conditions including perfectly conducting walls or vacuum surroundings, two cases relevant for many astrophysical, geophysical, and industrial flows. A spectral Fourier continuation method is used to accurately represent all fields and their spatial derivatives, allowing also for efficient solution of Poisson equations with different boundaries. A study of conducting flows at different Reynolds and Hartmann numbers, and with different boundary conditions, is presented to study convergence of the method and the accuracy of the solenoidal and boundary conditions.
\end{abstract}

\maketitle
\section{Introduction}
Numerical solutions to partial differential equations (PDEs) have been a cornerstone of engineering and physics research in the last decades, and their ubiquity has grown as computational power has increased. Numerical treatment is particularly relevant when dealing with problems modeled by non-linear PDEs, which are often impossible to treat analytically even for simple settings. The development of algorithms to yield improved numerical solutions, and to make a more efficient use of computational resources, remains to the present day an important research topic. In the particular case of fluid dynamics and plasma physics, if periodic boundary conditions can be used (e.g., when considering the bulk flow dynamics), employing Fourier representations is a well established method, as it tends to be optimal when considering the accuracy and computational efficiency of the numerical scheme \cite{Orszag1972,Pouquet1978}.

There is, nevertheless, a plethora of physical scenarios that cannot be successfully modeled employing PDEs with periodic boundary conditions. In those situations the zoology of numerical techniques is plentiful, and the method of choice is usually problem dependent \cite{Patera1984,Shan1991,Fornberg1998,Lele1992,Kwok2001,Canuto2006,Rosenberg2007,Julien2009,Fontana2018}. One strategy among many possible is to employ Fourier representations and adjust boundary conditions, most commonly, using either penalization methods \cite{Peskin1972,Wirth2005}, or periodic extensions of the non-periodic fields \cite{Boyd2002,Bruno2003}. However, the former tend to result in low order methods, whereas the latter can become prohibitively computationally expensive and hence, historically, Fourier-based methods have not seen much use for non-periodic problems, resulting in the use of other bases (in many cases more expensive to transform), or of lower-order methods for many applications.

In recent years, an efficient Fourier representation for fields with arbitrary boundary conditions was introduced, which uses Gram polynomials (sometimes referred to as discrete Chebyshev polynomials) to obtain high-order periodic extensions at a marginal computational cost, resulting in the so-called FC-Gram method \cite{Albin2011,Albin2012}. Besides yielding high order accuracy, this technique produces spatially dispersionless derivatives \cite{Bruno2014} --- in view of its reliance on Fourier expansions --- and hence phase speed propagation errors might arise solely as a result of the time stepping strategy. Another advantage of the FC-Gram method is that Poisson equations, which are common in some formulations of the incompressible hydrodynamic equations, can be easily and efficiently solved in bounded domains \cite{Fontana2020}. As a result, to the present day the FC-Gram method has been successfully employed in a wide range of PDE problems \cite{Albin2012,Amlani2016,Bruno2019,Fontana2020}.

Magnetohydrodynamic (MHD) equations, and related equations of electromagnetism and plasma physics, have many features in common with hydrodynamics but also their own complexities, and as a result have remained so far unexplored using FC-Gram methods. These non-linear equations, which model conducting fluids and the low frequency regime of plasmas \cite{Davidson2001}, are of major relevance to numerous research topics such as planetary and celestial magnetic fields \cite{Brandenburg2005,Jones2000,Roberts2000,Brun2004,Fan2014,Cattaneo2003}, stellar atmospheres \cite{Goldstein1995,Leamon2000}, non-mechanical pumps \cite{Kliman1979,Rodriguez2016}, and fusion research, to name a few. However, boundary conditions for the electromagnetic fields can be numerically harder to deal with, and the magnetic field must be kept solenoidal or otherwise measurable differences in the numerical solutions can be found \cite{Brackbill1980,Balsara2004}. Moreover, the MHD equations present a purely magnetic invariant, the magnetic helicity, whose definition involves the magnetic vector potential \cite{Matthaeus1982}. It is, therefore, also important for MHD research to be able to confidently rely on numerical methods which produce an accurate vector potential and, by proxy, are reliable for computing all the known ideal invariants of the system including the magnetic helicity. As a result, methods which evolve equations for the vector potential instead of the magnetic field are of particular relevance, as they have the benefits of explicitly enforcing the solenoidal condition on the magnetic field and of allowing direct computation of the magnetic helicity.

In this paper we present an algorithm (and a publicly available numerical code) for the solution of the incompressible MHD equations in the vector potential formulation, for a single non-periodic dimension along which FC-Gram representations are employed to evaluate derivatives. In particular, the highly relevant scenarios where the boundary conditions correspond to either perfectly conducting walls or vacuum surroundings are considered, as often used in simulations of the geodynamo or in industrial applications. Use of an explicit time integration scheme is proposed, as we will focus on the moderate and low diffusivity regime, where turbulent behaviour takes place and hence the Courant–Friedrichs–Lewy (CFL) constraint is dominated by field advection (i.e., by non-linearities). As indicated above, for simplicity, the method presented in this work is restricted to cases with a single non-periodic direction. It is however straightforward to extend the method presented here to additional non-periodic dimensions and to general non-rectangular domains, as well as to the compressible case and to other physical systems such as Hall-MHD flows, two-fluid plasma descriptions, or other MHD problems including systems with thermal convection or in rotating frames. Convergence of the method and the accuracy in satisfying the solenoidal and boundary conditions will be illustrated considering the Hartmann flow.

The main results are: (1) The method retains all the advantages of spectral methods for dealing with strongly non-linear systems, including being dispersionless and fast (as its based on FFTs). (2) Being a high-order method, and based on a high-order representation of the vector potential, it satisfies the solenoidal condition on the magnetic field down to an error of the order of floating point arithmetic limitations. (3) Solving Poisson equations (e.g., to impose the Coulomb gauge) is easy and well conditioned. (4) The proposed method can accommodate different types of boundary conditions for the magnetic field, and in particular, cases relevant for geophysical, astrophysical and industrial flows are explicitly treated. And finally, (5) a modified FC-Gram method allows for flexible and efficient numerical treatment of boundary conditions on second derivatives of the fields, and on boundary conditions prescribed by differential equations as in the case of Robin boundary conditions.

The paper is organized as follows. In \cref{sec:equations} we introduce the equations and geometry of interest. In \cref{sec:boundary} we derive appropriate boundary conditions for the vector potential. In \cref{sec:fc-gram}, an overview of the standard FC-Gram method is presented, and appropriate generalizations for the boundary conditions in question are discussed. In particular, we point the reader to \cref{sec:fc-robin} which introduces the modified FC-Gram approach for Robin boundary conditions --- without which the applicability of the FC-Gram method in this context would be compromised, in view of its ubiquity in electromagnetic problems and of its potentially high computing costs. Then, in \cref{sec:solver} a full time stepping algorithm is proposed for evolving the MHD equations. As an application, in \cref{sec:results} we consider the Hartmann flow scenario from a physical standpoint, while in \cref{sec:convergence} the numerical performance of the algorithm is evaluated. Finally, the conclusions are presented in \cref{sec:conclusions}.
\section{Governing equations}
\label{sec:equations}
The standard field formulation for the MHD approximation of an incompressible conducting fluid is encompassed in the set of equations
\begin{align}
	\label{equations:eq:continuity1}
	\bm \nabla \cdot \bm u &= 0,\\
	\label{equations:eq:navier-stokes1}
	\dpd{\bm u}{t} + (\bm{u} \cdot \bm \nabla ) u &= - \bm \nabla p + \bm j \times \bm b + \nu \nabla^2 \bm u, \\
	\label{equations:eq:magnetic-gauss1}
	\bm \nabla \cdot \bm b  &= 0, \\
	\label{equations:eq:induction}
	\dpd{\bm b}{t} &= \bm \nabla \times (\bm u \times \bm b) + \eta \nabla^2 b.
\end{align}
Here $\bm u, p, \bm b$, and $\bm j = \bm \nabla \times \bm b$ are the velocity, the pressure per unit mass density, the magnetic field, and the current density, respectively. The kinematic and magnetic diffusivities are $\nu$ and $\eta$, and the mass density is assumed to be homogeneous. The magnetic field $\bm b$ is written in velocity (Alfvénic) units. The magnetic permeability $\mu$ is equal to 1, a reasonable choice in most optical, geophysical, and astrophysical contexts. In this approximation the electric field $\bm E$ is reduced to a secondary role as a result of the quasineutrality hypothesis, although it can be recovered a posteriori from the non-relativistic Ohm's law
\begin{equation}
	\label{equations:eq:ohm}
	\eta\bm j = \bm E + \bm u \times \bm b.
\end{equation}

Equation (\ref{equations:eq:magnetic-gauss1}) implies that the magnetic field is always solenoidal, and imposes a condition on advection and stretching of magnetic field lines \cite{Davidson2001}. In low-order methods, or in approximate methods, it has been shown that solutions are sensitive to the way this condition is enforced, or on how well it is satisfied \cite{Brackbill1980}. In particular, measurable differences in physical quantities were reported for different methods \cite{Balsara2004}. When treating problems analytically, the condition $\bm \nabla \cdot \bm b = 0$ can be enforced by representing the magnetic field via the vector potential $\bm a$ which is defined by the relation $\bm \nabla \times \bm a = \bm b$. This latter expression, however, does not define a unique vector potential, as adding a gradient to $\bm a$ results in the same magnetic field. To properly define a vector potential both $\bm \nabla \cdot \bm a$ and boundary conditions compatible with those for the magnetic field must be prescribed. In non-relativistic MHD, the most common choice is to use the Coulomb gauge, $\bm \nabla \cdot \bm a = 0$.

Besides guaranteeing $\bm \nabla \cdot \bm b = 0$, the vector potential is also important to define the magnetic helicity density, $h^m = \bm a \cdot \bm b$. The volumetric integral of this quantity measures the topological complexity of magnetic field lines, and is the only purely magnetic ideal invariant of the MHD equations, the other two known invariants involving also the velocity field being the volumetric integral of the total energy density, $e = {\bm u}^2 + {\bm b}^2$, and the volumetric integral of the cross helicity density, $h^c = \bm u \cdot \bm b$ \cite{Davidson2001,Galtier2016}. It is therefore desirable in the numerical study of MHD to ensure an accurate computation of the magnetic helicity, for which the vector potential is needed.

If the magnetic field is represented via a vector potential in the Coulomb gauge, integrating \cref{equations:eq:induction} results in the following set of equations for the MHD approximation of an incompressible flow in terms of $\bm a$,
\begin{align}
	\label{equations:eq:continuity2}
	\bm \nabla \cdot \bm u &= 0,\\
	\label{equations:eq:navier-stokes2}
	\dpd{\bm u}{t} + (\bm u \cdot \bm \nabla) \bm u &= - \nabla p - \nabla^2 \bm a \times (\bm \nabla \times \bm a) + \nu \nabla^2 \bm u, \\
	\label{equations:eq:coulomb}
	\bm \nabla \cdot \bm a &= 0 \\
	\label{equations:eq:induction-a}
	\dpd{\bm a}{t} &= \bm u \times \left( \bm \nabla \times \bm a \right) + \eta \nabla^2 \bm a - \bm \nabla \phi.
\end{align}
where we used $\bm j=-\nabla^2 \bm a$. The scalar potential $\phi$ results from the integration of \cref{equations:eq:induction}, and must fulfill the gauge condition $\bm \nabla \cdot \bm a = 0$. A close look at \cref{equations:eq:ohm} reveals that $\phi$ is no other than the standard electric scalar potential, recovering the usual expression for the electric field
\begin{equation}
	\bm E = -\bm \nabla \phi - \dpd{\bm a}{t}.
	\label{equations:eq:electric-field}
\end{equation}
It should also be noted that, under this standard formulation of the incompressible MHD equations, both the pressure and the electric potential must fulfill seldom Poisson equations, since taking the divergence of \cref{equations:eq:navier-stokes2,equations:eq:induction-a} and considering the solenoidality condition yields
\begin{align}
	\label{equations:eq:poisson-p}
	\nabla^2 p    &= - \bm{\nabla} \cdot \left[ \left(\bm{u} \cdot \bm{\nabla}\right) {\bm u}\right], \\
 	\label{equations:eq:poisson-phi}
	\nabla^2 \phi &=   \bm{\nabla} \cdot \left[ \bm u \times \left(\bm \nabla \times \bm a \right)
	 \right].
\end{align}

\begin{figure*}
	\centering
	\includegraphics[width=.9\linewidth, keepaspectratio]{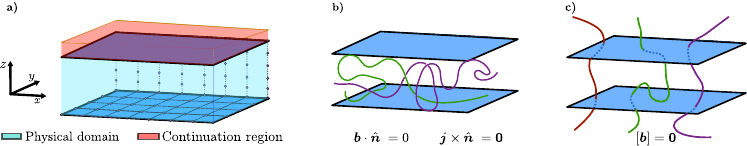}
	\caption{\textbf{a)}:Illustration of the explored geometry, a cuboid domain, discretized employing a uniform grid in each of $\hat{\bm x}$, $\hat{\bm y}$, and $\hat{\bm z}$ directions. For simplicity, the only non-periodic direction is $\hat{\bm z}$. The actual physical domain is represented in light blue, whereas the spaced used to compute the FC-Gram continuations is shaded in red. \textbf{b)} Sketch of the magnetic field lines inside the domain when considering perfectly conducting boundary conditions, together with the appropriate boundary conditions. \textbf{c)} Sketch of the magnetic field inside and outside the domain occupied by the fluid for the case of vacuum boundary conditions.}
	\label{equations:fig:geometry}
\end{figure*}

The numerical method presented in this work can compute solutions for \cref{equations:eq:navier-stokes2,equations:eq:induction-a,equations:eq:poisson-p,equations:eq:poisson-phi} after boundary conditions for the physical fields are provided. We will consider a $(0,0,0) \times (L_x, L_y, L_z)$ cuboid domain, and a uniform spatial discretization in each direction consisting of $N_x \times N_y \times N_z$ gridpoints (see \cref{equations:fig:geometry}).

\section{Boundary conditions for the vector and scalar potentials}
\label{sec:boundary}
\subsection{Perfectly conducting boundary conditions}
A relevant scenario for astrophysical and geophysical applications is that of a magnetofluid that is confined by a perfectly conducting medium. This is the situation encountered, e.g., in many simulations of the geodynamo when treating the boundary with the inner core \cite{Roberts2000}. In this case the electric field must vanish inside the conductor, and there cannot exist a time-varying magnetic field in its interior. Considering now a stationary conductor ($\bm u = \bm 0$), the continuity of the normal component of the magnetic field, and Ohm's law, the magnetic boundary conditions that the system must fulfill are
\begin{align}
	\label{boundary:eq:j}
	\bm j|_{\partial \Omega} \times \hat{\bm n} &= \bm 0,\\
	\label{boundary:eq:d_t-b}
	\partial_t \bm b|_{\partial \Omega} \cdot \hat{\bm n} &= \bm 0,
\end{align}
where $\partial \Omega$ denotes the boundary surface and $\hat{\bm n}$ is a normal unit vector at $\partial \Omega$. Note that if $\bm b \cdot \hat{\bm n} = 0$ at $t=0$, then $\bm b \cdot \hat{\bm n} = 0 \quad \forall t>0$. For simplicity we consider this latter scenario from here onwards.

Assuming now periodic boundary conditions in the $\hat{\bm x}$ and $\hat{\bm y}$ directions and the presence of a conducting wall at $z=0$, it follows that a vector potential satisfying the Coulomb gauge constraint, $\bm a|_{z=0} \times \hat{\bm z}=0$, and $(\partial^2 \bm a / \partial z^2)|_{z=0} \times \hat{\bm z}=0$ at $t=0$ verifies the boundary conditions given by \cref{boundary:eq:j,boundary:eq:d_t-b}. Note that the normal component of the vector potential, $a_\perp = a_z$, is not directly involved in the determination of any of the physical boundary conditions. For this to be the case, however, the Coulomb gauge must be maintained at all times and thus it is reasonable for $a_z$ at the boundary to explicitly enforce the interface gauge condition, resulting in the set of conditions
\begin{align}
	\label{boundary:eq:bc-perfect-a}
	\bm a|_{z=0} \times \bm z &= 0, \\
	\label{boundary:eq:bc-perfect-da}
	\left. \dpd[2]{\bm a}{z} \right |_{z=0} \times \hat{\bm z} &= \bm 0, \\
	\label{boundary:eq:bc-perfect-div}
	\left. \dpd{\bm a}{z} \right|_{z=0} \cdot \hat{\bm z} &= 0, \\
	\label{boundary:eq:bc-perfect-phi}
	\phi|_{z=0} &= 0,
\end{align}
which guarantee the physical boundary conditions are maintained at all times. Even more, choosing $\phi = 0$ in the boundary is in full agreement with the fact that the perfect conductor must be an equipotential surface. Although the boundary conditions above appear at first sight to overdetermine the problem, it should be noted that some choices are made using the gauge freedom, and that if they are enforced exactly the dynamical equations imply that $\bm a_\parallel|_{z=0} = \bm 0$ acts only as an initial condition (with the subindex indicating the components parallel to the boundary), and that together with $(\partial^2 \bm a_\parallel/\partial z^2)|_{z=0} = \bm 0$, which is the physical boundary condition, suffice to enforce the conditions on the magnetic field and the current density.

\subsection{Vacuum boundary conditions}
\label{sec:boundary:vacuum-theoretical}
Another relevant astrophysical and geophysical scenario is that of a magnetofluid which is surrounded by vacuum. As mentioned before, we assume for this work that in the magnetofluid $\mu=1$, and hence the magnetic field at the interface must be continuous, that is
\begin{equation}
	\label{eq:b-cont}
	\left[ \bm b \right]_{\partial \Omega} = \bm 0 \qquad \Longrightarrow \qquad \bm b^\I |_{\partial \Omega} = \bm b^\II|_{\partial \Omega},
\end{equation}
where $^\I$ and $^\II$ denote quantities in the fluid and vacuum media respectively, $\partial \Omega$ is the boundary between them, and $[\bm q]_S$ denotes the jump in field $\bm q$ across $S$. Expanding both fields in terms of the respective vector potentials $\bm b^i = \bm \nabla \times \bm a^i$ it is straightforward to get the corresponding bulk equations for $\bm a^\II$ in the vacuum domain, as the current density must necessarily vanish there. Due to the absence of electric charge, the vacuum electric potential $\phi^\II$ must be harmonic. The respective equations are
\begin{align}
	\nabla^2 \phi^\II &= 0,\\
	\nabla^2 \bm a^\II &= 0,\\
	\bm \nabla \cdot \bm a^\II &= 0,
\end{align}
where, as before, the Coulomb gauge was chosen for the vector potential.
	
The physical boundary condition, \cref{eq:b-cont}, implies that $\bm a$ must obey
\begin{equation}
	\label{eq:a-curl}
	(\bm \nabla \times \bm a^\I)_{\partial \Omega} = (\bm \nabla \times \bm a^\II)_{\partial \Omega},
\end{equation}
whereas regularity conditions on $\bm b$ require $\bm a$ to be continuous at the boundary, that is, $\bm a^\I = \bm a^\II$.
	
To address the physical boundary conditions for the scalar potential, we start by formulating appropriate boundary conditions for the electric field, which are
\begin{align}
	\bm E^\I \times \hat{\bm n}|_{\partial \Omega} = \bm E^\II \times \hat{\bm n}|_{\partial \Omega}, \\
	\bm E^\I \cdot \hat{\bm n}|_{\partial \Omega} = \frac{\epsilon^\I}{\epsilon ^\II}\bm E^\II \cdot \hat{\bm n}|_{\partial \Omega},
\end{align}
with $\epsilon^i$ the electrical permittivity of medium $i$, and $\hat{\bm n}$ a normal unit vector pointing from $\I$ towards $\II$. For a sizeable class of conducting fluids the permittivity can be approximated by that of vacuum, and hence the normal component of the electric field is continuous in the boundary, which expressed in terms of the potentials reads
\begin{equation}
	\left(\dpd{\bm a^\I}{t} \cdot \hat{\bm n} + \dpd{\phi^\I}{n} \right)_{\partial \Omega} = 	\left(\dpd{\bm a^\II}{t} \cdot \hat{\bm n} + \dpd{\phi^\II}{n} \right)_{\partial \Omega}.
\end{equation} 
Considering the continuity of the vector potential across the boundary, the latter equation reduces simply to the continuity of the normal derivative of the scalar potential. Analyzing a similar equation for the parallel electric field leads to the continuity of the electric potential itself.

For the case of periodic boundary conditions in $\hat{\bm x}$ and $\hat{\bm y}$, assuming vacuum is present in the semispace $z>0$ and considering only bounded solutions for $z \to \infty$, the harmonic equations for $\bm a^\II$ and $\phi^\II$ admit the solutions
\begin{align}
	\bm a^\II (x,y,z) &= \sum_{m=-\infty}^{\infty} \sum_{n=-\infty}^{\infty} \hat{\bm a}^\II_{nm} e^{i(k^x_n x + k^y_m y) -\gamma_{nm} z},\\
	\phi^\II (x,y,z) &= \sum_{m=-\infty}^{\infty} \sum_{n=-\infty}^{\infty} \hat{\phi}^\II_{nm} e^{i(k^x_n x + k^y_m y) -\gamma_{nm} z} ,
\end{align}
where $k^x_n$ (resp.~$k^y_m$) is the $n$-th (resp.~$m$-th) wavenumber in the $\hat{\bm x}$ (resp.~$\hat{\bm y}$) direction and ${\gamma_{nm} = [(k^x_n)^2 + (k^y_m)^2]^{1/2}}$. For the vector potential, the condition
\begin{equation}
	a^{z\II}_{nm} = \frac{i}{\gamma_{mn}} ( k^x_n a^{x\II}_{nm} + k^y_m a^{y\II}_{nm}),
\end{equation}
corresponds to enforcement of the Coulomb gauge.

On the basis of this solution boundary conditions for the vector potential $\bm a^\I$ can be obtained easily on the base of its continuity, the physical boundary conditions in \cref{eq:a-curl} and the gauge continuity across the boundary, leading to the set of conditions
\begin{equation}
	\left. \left( \dod{\bm a^{\I}_{nm}}{z} + \gamma_{nm} \bm a^{\I} \right ) \right|_{z=0} = 0,
	\label{eq_robin_a}
\end{equation}
which represents an homogeneous Robin boundary condition for each wavenumber component of the vector potential in the fluid medium. In a similar way, the continuity of both $\partial_z \phi$ and $\phi$ itself leads to a fully analogous Robin condition
	\begin{equation}
		\left. \left( \dod{\phi^{\I}_{nm}}{z} + \gamma_{nm} \phi^{\I} \right ) \right|_{z=0} = 0.
	\end{equation}
It is worth emphasizing that these boundary conditions assume vacuum is present in the semispace $z>0$. Naturally, if vacuum is instead in the semispace $z<0$, the corresponding harmonic solution has a factor $e^{\gamma_{nm} z}$ (note the absence of the minus sign), and the resulting boundary conditions for this case would be
\begin{equation}
	\left. \left( \dod{}{z} - \gamma_{nm} \right) \right|_{z=0} = 0,
\end{equation}
for $\phi$ and each cartesian component of $\bm a$.

Robin boundary conditions are common in many electromagnetic problems, and as a result, we now present a generalization of the FC-Gram method to deal with such conditions, as well as with the boundary conditions discussed before for the case of a perfect conductor.
\section{Generalization of FC-Gram to boundary conditions in MHD}
\label{sec:fc-gram}
\subsection{FC-Gram with Dirichlet boundary conditions}
In order to have an accurate but computationally efficient representation of the fields and their derivatives, all the relevant variables are projected onto a Fourier representation basis. As both the $\hat{\bm x}$ and $\hat{\bm y}$ directions have periodic boundary conditions, the transformation to the wavenumber domain can be directly computed via standard FFT operations. That is not the case, however, for the non-periodic $z$ dimension, as the well known Gibbs ringing phenomenon \cite{Bocher1906} would severely degrade accuracy in the representation of the fields' derivatives. To circumvent this limitation we employ a computationally efficient continuation methodology, known as FC-Gram, first introduced in \cite{Lyon2010a,Albin2011} and recently utilized in \cite{Fontana2020} for a hydrodynamic Navier-Stokes solver.

\begin{figure}
	\centering
	\includegraphics[width=\linewidth, keepaspectratio]{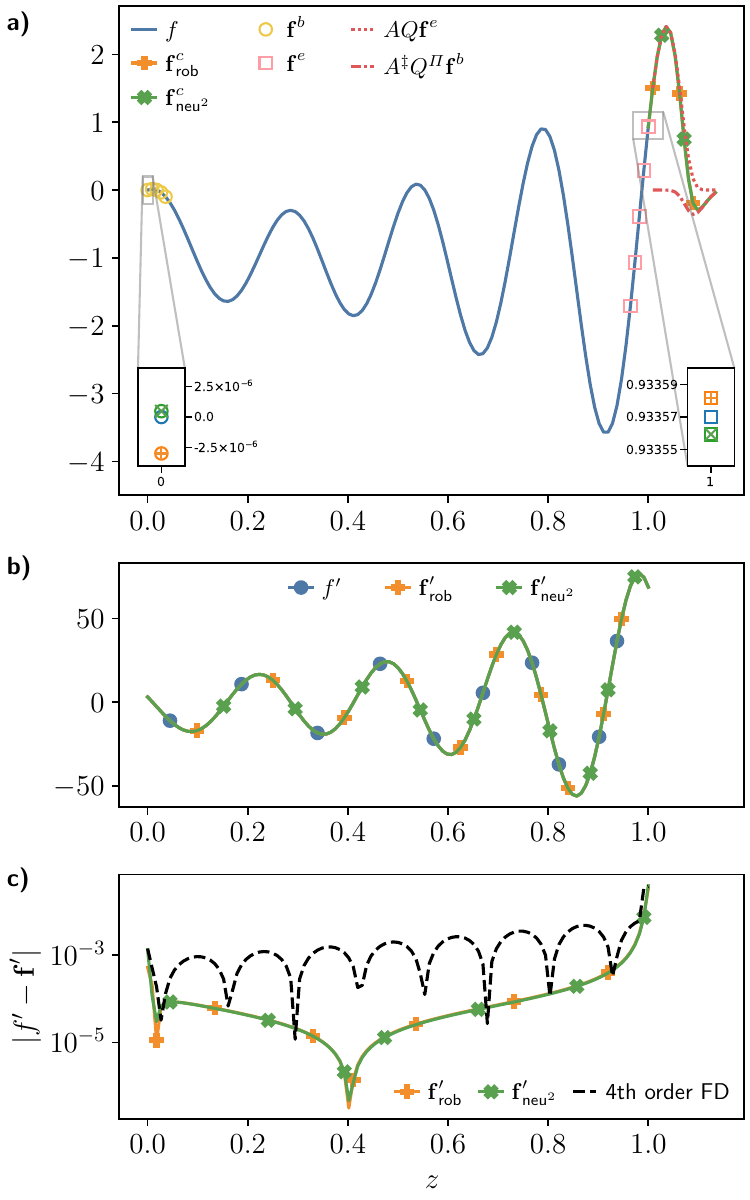}
	\caption{Example of a 4\textsuperscript{th} order ($d=5$) FC-Gram procedure applied to the function $f(z)=J_0(25z)e^{3z}-1$ for $N=113$ grid points in the $[0, 1]$ domain and $C=15$ continuation points. Panel \textbf{a)} shows the function over the discrete grid (in blue) together with the periodic continuations computed using the second normal derivative $\bm{\mathrm{f}}^c_{\text{neu}^2}$ (in green with $\times$ markers) and the Robin condition $f'+ 5f$, $\bm{\mathrm{f}}^c_\text{rob}$ (in orange with + markers), as the boundary condition at both endpoints. Also displayed are the matching values used to perform the continuations, $\bm{\mathrm{f}}^b$ (in yellow circles) and $\bm{\mathrm{f}}^e$ (in pink squares), as well as the blend to zero continuations associated to $\bm{\mathrm{f}}^e$ ($AQ \bm{\mathrm{f}}^e$) with a red dotted line, and $\bm{\mathrm{f}}^b$ ($A^\ddagger Q^{\varPi} \bm{\mathrm{f}}^b$) with a red dashed-dotted line. Insets show $f(0)$ and $f(1)$ together with the values reconstructed from the boundary conditions. Panel \textbf{b)} shows the exact derivative $f'$ (in blue using circle markers) together with the ones computed spectrally from $\bm{\mathrm{f}}^c_{\text{neu}^2}$ (in green with $\times$ markers) and $\bm{\mathrm{f}}^c_\text{rob}$ (in orange with + markers) over the $[0, 1]$ grid. Panel \textbf{c)} shows the absolute error in the derivative estimation for both $\bm{\mathrm{f}}^c_{\text{neu}^2}$ (in green with $\times$ markers) and $\bm{\mathrm{f}}^c_\text{rob}$ (in orange with + markers). For comparison, a same order finite difference error is also exhibited (with a dashed black line). Note that for display purposes marker spacing is larger than grid spacing in the whole figure, except for the quantities $\bm{\mathrm{f}}^b$ and $\bm{\mathrm{f}}^e$ for which grid and marker spacing are the same.}
	\label{fc-gram:fig:fc}
\end{figure}

The idea behind a Fourier Continuation (FC) technique can be summarized as follows for the one dimensional case. Let $f(z)$ be a non-periodic function defined over the discrete grid $z_i=i \Delta z$, with $\Delta z$ the (uniform) grid spacing and $i=0,\ \hdots{} \,N-1$, resulting in $\bm{\mathrm{f}} = \big[f(z_0), \hdots , f(z_{N-1}) \big]$. The method generates an efficient continuation spanning the domain $z_N,\ \hdots \ , z_{N+C-1}$, where $C$ is the number of points used for the extension operation and is a parameter of the method. The resulting quantities $\bm{\mathrm{f}}^c = \big[ f(z_{N}),\ \hdots \ , f(z_{N+C-1})\big]$ can then be appended to the values of the function over the original grid, as depicted in \cref{fc-gram:fig:fc}, and a Fourier representation can be computed for $\bm{\mathrm{f}} \cup \bm{\mathrm{f}}^c$ over the interval $[z_0, z_{N+C})$. Any derivatives estimated in the wavenumber domain can then be inverse Fourier transformed to get accurate representations over the original $[z_0, z_{N-1}]$ grid.

To obtain $\bm{\mathrm{f}}^c$ in a computationally efficient manner, the FC-Gram method uses only information near the boundary to estimate $\bm{\mathrm{f}}^c$. To this end, lets first consider the simpler case where we have $d$ values at the end of the domain $\bm{\mathrm{f}}^e_\text{dir} = \big[f(z_{N-d}), \ \hdots \ , f(z_{N-1})\big]$, and we wish to compute a continuation that smoothly transitions from $f(z_{N-1})$ to $0$. The subindex ``dir" is used to denote a Dirichlet boundary condition on the endpoint, i.e., that $f(z_{N-1})$ is known. The resulting extension will hence have a $d$-th order of approximation at the boundary. Note $d$ (the number of points near the boundary used to compute the continuation) is an additional parameter of the FC-Gram technique.

Fixing $d$ defines a set of discrete polynomials orthogonal with respect to the inner product 
\begin{equation}
	\braket{\bm{\mathrm{g}}}{\bm{\mathrm{h}}} = \sum_{i=0}^{d-1} g(x_i) h(x_i),
\end{equation}
i.e., the Gram polynomials. Note that $x_i$ is used in this equation as the inner product definition is unrelated to the grid $z_i$ introduced before. An operator of length $C$ that smoothly transitions each Gram polynomial to zero (i.e., a blend to zero operator) can be numerically obtained, resulting in the $C\times d$ blend-to-zero matrix $A_\text{dir}$ \cite{Amlani2016}. The subindex ``$\text{dir}$" again denotes the fact that this matrix is for Dirichlet boundary conditions. Similarly, an operator that projects arbitrary function values at $d$ grid points onto each of the Gram polynomials can be obtained, resulting in a certain matrix $Q_\text{dir}$ (see \cite{Amlani2016,Fontana2020} for more details).

The computation of operators $A_\text{dir}$ and $Q_\text{dir}$ must be performed utilizing arbitrary precision linear algebra routines, as it requires decomposing ill-conditioned matrices. However, it is worth noting that these computations, which require only a few minutes in a single modern CPU core (although its computation can be parallelized), must be performed only one time for a given selection of $C$ and $d$, with the resulting operators being utilized henceforth. Additional details about obtaining these operators can be found in \cite{Lyon2010a,Albin2011}, and also in \cite{Fontana2020} for a case with a very similar scope to the one proposed in this work.

With the $A_\text{dir}$ and $Q_\text{dir}$ operators at hand, finding a continuation that smoothly transitions $\bm{\mathrm{f}}$ to $0$ is straightforward, requiring only a projection of the boundary end values $\bm{\mathrm{f}^}e_\text{dir}$ onto the Gram space, and blending each of those polynomials to zero. The resulting smooth extension to zero $\tilde{\bm{\mathrm{f}}}^c$ is computed as
\begin{equation}
	\tilde{\bm{\mathrm{f}}}^c = A_\text{dir} Q_\text{dir} \bm{\mathrm{f}}^e_\text{dir}.
\end{equation}

It is now possible to consider the original problem in which we want a smooth transition from $\bm{\mathrm{f}}^e_\text{dir}$ to the $d$ values at the beginning of the domain $\bm{\mathrm{f}}^b_\text{dir} = \big[f(z_0), \hdots , f(z_{d-1})\big]$, to get a periodic extension. This can be easily solved now by adding to $\tilde{\bm{\mathrm{f}}}^c$ a set of values which smoothly transition from 0 to $f(z_0)$. A suitable continuation which transitions from $f(z_{N-1})$ to $f(z_0)$ and which has $d-1$ smooth derivatives at the endpoints can thus be constructed as
\begin{equation}
	\label{fc-gram:eq:continuation}
	\bm {\mathrm{f}}^c = A_\text{dir} Q_\text{dir} \bm{\mathrm{f}}^e_\text{dir} + A^\ddagger_\text{dir} Q^{\varPi}_\text{dir} \bm{\mathrm{f}}^b_\text{dir},
\end{equation}
where $^\ddagger$ and $^{\varPi}$ denote the row-reversing and column-reversing operations, respectively. The underlying concept in \cref{fc-gram:eq:continuation}, which can be easily appreciated in \cref{fc-gram:fig:fc}, is the superposition of values which smoothly transition from $f(z_{N-1})$ to zero (given by $A_\text{dir} Q_\text{dir} \bm{\mathrm{f}}^e_\text{dir}$), and from zero to $f(z_0)$ (prescribed by $A^\ddagger_\text{dir} Q^{\varPi}_\text{dir} \bm{\mathrm{f}}_\text{dir}^b$), resulting in the overall smooth periodic extension.

It is worth pointing out that, as mentioned earlier, $A_\text{dir}$ and $Q_\text{dir}$ depend only on the choice of $d$ and $C$, and not on the data to be continued $\bm{\mathrm{f}}^b_\text{dir}$, $\bm{\mathrm{f}}^e_\text{dir}$. This means that the ill-conditioning usually found when trying to compute suitable periodic extensions is encapsulated in the estimation of those matrices.  This is specially important for 3D problems, as the fastest way to obtain the values on the extended volume (as depicted in \cref{equations:fig:geometry}) is to simply apply \cref{fc-gram:eq:continuation} to each $xy$-line (i.e., by computing $N_x \times N_y$ 1D continuations). Another detail worth noting is that the same $A_\text{dir}$ and $Q_\text{dir}$ operators can accurately continue both the real and imaginary parts in case the vectors $\bm{\mathrm{f}}^b_\text{dir}$, $\bm{\mathrm{f}}^e_\text{dir}$ are complex valued, allowing to commute FFTs and FC-Gram operations (which is important for efficient parallelization of the method, see \cite{Mininni2011,Fontana2020}). It is also worth mentioning that, depending on the problem at hand, great accuracy is obtained when considering $d$ in the range of $4$ to $10$ and $C$ in the range of $15$ to $33$. This has the added benefit that consequently both $A_\text{dir}$ and $Q_\text{dir}$ fit easily in the L1 cache of modern CPU cores.

Within the context of PDE solvers, the FC-Gram picture described above is useful when the function values are known everywhere, that is, both at every interior point as well as at the boundary. For the case of having the normal derivative prescribed at one end of the domain, a solution within the framework of FC-Gram which is both accurate and efficient was introduced in \cite{Amlani2016}. A treatment for the case in which second normal derivatives are prescribed at the boundary is introduced next in \cref{sec:fc-second-normal}. A straightforward extension of these methods to Robin boundary conditions is not efficient, as in electromagnetism a large number of such conditions may arise simultaneously with different Robin coupling parameters for each mode (as, e.g., in Eq.~\ref{eq_robin_a}). Thus, a new FC-Gram algorithm for the Robin problem that allows quick computation for different Robin coupling parameters is presented in \cref{sec:fc-robin}.

\subsection{Modified FC-Gram method for prescribing the second normal derivative}
\label{sec:fc-second-normal}
We are once again interested in finding a suitable blend to zero procedure using the information at the end of the domain, but now for the case where $d-1$ values at the interior points and the second derivative at the last grid point are known. Once found, that blend to zero procedure can be easily employed to compute a periodic extension, as it was previously discussed. Therefore, in a similar way to the ideas introduced for the Dirichlet case, one possible strategy to deal with this problem within the framework of the FC-Gram method is to project the last $d-1$ interior values plus the prescribed second normal derivative onto a space of modified Gram polynomials, ones which are orthogonal with respect to the inner product
\begin{equation}
	\braket{\bm{\mathrm{g}}}{\bm{\mathrm{h}}} = g''(x_{d-1}) h''(x_{d-1}) + \sum_{i=0}^{d-2} g(x_i) h(x_i).
\end{equation}

Analogously to the procedure described in \cite{Albin2011,Amlani2016}, given an arbitrary uniform grid $x_0, \hdots , x_{d-1}$ with spacing $\Delta x$, a set of such modified Gram polynomials can be obtained employing the modified Vandermonde matrix
\begin{equation}
	P_{\text{neu}^2} = \begin{pmatrix}
		1 		& x_0 		& x_0^2 	& \hdots 	& x_0^{d-1}		\\
		1 		& x_1 		& x_1^2 	& \hdots 	& x_1^{d-1}		\\
		\vdots 	&\vdots 	& \vdots 	& \ddots 	& \vdots 		\\
		1		& x_{d-2}   & x_{d-2}^2 & \hdots	& x_{d-2}^{d-1}	\\
		0		& 0			& 2 		& \hdots    & (d-1)(d-2)x_{d-1}^{d-3}
	\end{pmatrix},
\end{equation}
via its $QR$ decomposition $P_{\text{neu}^2} = Q_{\text{neu}^2} R_{\text{neu}^2}$ (the subindex ``neu$^2$" here stands for the case with Neumann conditions on the second derivative). It is then possible to proceed and construct blend-to-zero operators for this set of polynomials in an analogous way to the Dirichlet case, obtaining an operator $A_{\text{neu}^2}$. From these matrices a blend-to-zero continuation can be computed given interior values and the respective second derivative, $\bm{\mathrm{n}} = \big [ f(x_0), \ \hdots \ , f(x_{d-2}), f''(x_{d-1}) \big ]$, directly as $A_{\text{neu}^2} Q_{\text{neu}^2} \bm{\mathrm{n}}$.

One comparably accurate and more computationally convenient procedure is to obtain the function value at $x_{d-1}$ employing the known function values at $x_0, \hdots, x_{d-2}$ and the second derivative at the endpoint. Once $f(x_{d-1})$ is known, continuation values can be computed using the Dirichlet blend to zero operator $A$ by means of \cref{fc-gram:eq:continuation}. For this purpose it is useful to first introduce the standard Vandermonde matrix 
\begin{equation}
	P_\text{dir} = \begin{pmatrix}
		1 		& x_0 		& x_0^2 				& \hdots 	& x_0^{d-1}		\\
		1 		& x_1 		& x_1^2 				& \hdots 	& x_1^{d-1}		\\
		\vdots 	&\vdots 	& \vdots 				& \ddots 	& \vdots 		\\
		1		& x_{d-2}   & x_{d-2}^2 			& \hdots  	& x_{d-2}^{d-1}	\\
		1		& x_{d-1}   & x_{d-1}^2 			& \hdots  	& x_{d-1}^{d-1}	\\
	\end{pmatrix},
\end{equation}
with an associated QR factorization $P_\text{dir}=Q_\text{dir} R_\text{dir}$. The corresponding function end value can be obtained from the relations
\begin{align}
	P_{\text{neu}^2} \bm{c} &= \bm{\mathrm{n}}^T, \\
	P \bm{c} &= \bm{\mathrm{d}}^T,
\end{align}
with $\bm c$ the coefficient of each monomial term on the respective Gram basis and $\bm{\mathrm{d}} = \big [ f(x_0),\ \hdots \ , f(x_{d-2}), f(x_{d-1}) \big] $. Solving for $\bm c$ using the $Q R$ factorizations of the respective $P$ matrices results in
\begin{equation}
	Q_\text{dir} R_\text{dir} R_{\text{neu}^2}^{-1} Q_{\text{neu}^2}^T \bm{\mathrm{n}}^T = \bm{\mathrm{d}}^T,
\end{equation}
and hence the end value $f(x_{d-1})$ is given by
\begin{equation}
	f(x_{d-1}) = \tilde{\bm{q}}_{\text{neu}^2} \bm{\mathrm{n}}^T,
\end{equation}
where $\tilde {\bm{q}}_{\text{neu}^2}$ is the last row of $Q_\text{dir} R_\text{dir} R_{\text{neu}^2}^{-1} Q_{\text{neu}^2}^T$, that is $(\tilde{q}_{\text{neu}^2})_i = (Q_\text{dir} R_\text{dir} R_{\text{neu}^2}^{-1} Q_{\text{neu}^2}^T)_{d-1, i}$.

It should be noted that the order of accuracy in the reconstruction of $f(x_{d-1})$ is still $d$, independently of the fact that a second derivative is prescribed instead of the value of $f$ at $x_d$. Thus, if the reconstruction is performed utilizing $d_{\text{neu}^2}$ values at the end of the domain ($d_{\text{neu}^2}-1$ function values and the prescribed second derivative) after which a $C\times d_\text{dir}$ Dirichlet blend to zero operator $A_\text{dir}$ is used to compute continuation values, the choice $d_{\text{neu}^2} = d_\text{dir}$ should be favored to have a consistent order in the approximation of all quantities at the boundaries.

To see that prescribing $d-1$ points plus some derivative (resulting in $d$ data points) yields the aforementioned order for the approximation, both a quick argument and a more formal proof, can be given. As a first qualitative argument, let's assume we know the value of a function $f(x)$ at $0$ with an error of order $h^d$, and its first derivative with error $h^{d-1}$. Then, replacing in its Taylor expansion, $f(x) \approx f(0) + f'(0) h$, it is straightforward to see the error in both terms is of order $h^d$. More formally, we can use Lagrange's mean value theorem to see that there are $d-2$ intermediate points where the derivatives of the difference between the function and an interpolating polynomial vanish. Thus, the derivative of this polynomial is an interpolating polynomial of the derivative of the function on a set of $d-1$ points. Then the theorem that gives the polynomial interpolation error can be applied, and it will give order $h^{d-1}$. Integrating between $0$ and $(d-1) h$ gives the error between the polynomial and the function, giving an error of order $h^d$. Similar arguments can be written for the case in which the second derivative is prescribed, or for the case of Robin boundary conditions.

Finally, it is worth pointing out that in the Dirichlet case operators $A_\text{dir}$ and $Q_\text{dir}$ could be used for any grid, irrespective of the grid spacing used for generating them. However, when the second derivative is prescribed, care should be taken to correctly account for the difference in the spacings when utilizing $\tilde {\bm{q}}_{\text{neu}^2}$. Returning now to the original grid $z_0, \ \hdots \ , z_{N-1}$ with spacing $\Delta z$, and considering that a $\tilde {\bm{q}}_{\text{neu}^2}$ operator is available and was computed using a spacing $\Delta x$, the value $f(z_{N-1})$ can be computed from $f''(z_{N-1})$ by simple application of the chain rule, that is
\begin{equation}
	f(z_{N-1}) = \tilde{\bm{q}}_{\text{neu}^2} \bm{\mathrm{f}}_{\text{neu}^2}^T,
\end{equation}
with
\begin{equation}
	\bm{\mathrm{f}}_{\text{neu}^2} = \left[ f(z_{N-d}), \ \hdots \ , f(z_{N-2}), f''(z_{N-1}) (\Delta z/\Delta x)^2 \right].
\end{equation}

\subsection{Modified FC-Gram method for Robin boundary conditions}
\label{sec:fc-robin}
When faced with Robin boundary conditions the function values $f({z_i})$ at every interior point are known and the value of a certain linear combination between the function and its derivative at one end of the domain is prescribed, that is
\begin{equation}
	f'(z_{N-1}) + \lambda f(z_{N-1}) = g.
\end{equation}

The first procedure discussed in the preceding subsection can be also applied for this case. That is, from the discrete inner product
\begin{multline}
	\braket{\bm{\mathrm{g}}}{\bm{\mathrm{h}}} = \left[g'(x_{d-1}) + \lambda g(x_{d-1}) \right] \left[h'(x_{d-1}) + \lambda h(x_{d-1}) \right] +\\
		+ \sum_{i=0}^{d-2} g(x_i) h(x_i),
\end{multline}
a set of modified Gram polynomials is defined and a related projector $Q_\text{rob}$ and blend to zero operator $A_\text{rob}$ can be obtained from the following modified Vandermonde matrix
\begin{equation}
	P_\text{rob} = \begin{pmatrix}
		1 		& x_0 				& \hdots 	& x_0^{d-1}		\\
		1 		& x_1 				& \hdots 	& x_1^{d-1}		\\
		\vdots 	&\vdots 			& \ddots 	& \vdots 		\\
		1		& x_{d-2}   		& \hdots  	& x_{d-2}^{d-1}	\\
		1		& 1+\lambda x_{d-1}	& \hdots    & (d-1)x_{d-1}^{d-2} + \lambda x_{d-1}^{d-1}
	\end{pmatrix}.
\end{equation}
The $C$ blend to zero values can then be obtained from the vector ${\bm{\mathrm{r}} = \big [ f(x_0), \hdots , f(x_{d-2}), f'(x_{d-1}) + \lambda f(x_{d-1}) \big ]}$ as $A_\text{rob} Q_\text{rob} \bm{\mathrm{r}}$.

It is also possible, and more convenient, to evaluate the solution value at the endpoint $x_{d-1}$ employing the known values at $x_0, \hdots, x_{d-2}$ and the boundary condition at $x_{d-1}$. Once again, this can be attained from the relations
\begin{align}
	\label{eq:Prob-c}
	P_\text{rob} \bm{c} &= \bm{\mathrm{r}}^T, \\
	\label{eq:P-c}
	P_\text{dir} \bm{c} &= \bm{\mathrm{d}}^T,
\end{align}
which has the solution 
\begin{equation}
	Q_\text{dir} R_\text{dir} R_\text{rob}^{-1} Q_\text{rob}^T \bm{\mathrm{r}}^T = \bm{\mathrm{d}}^T,
\end{equation}
where $R_\text{rob}$ is the upper triangular matrix resulting from the $QR$ factorization of $P_\text{rob}$. The end value $f(x_{d-1})$ is then given by
\begin{equation}
	f(x_{d-1}) = \tilde{\bm{q}}_\text{rob} \bm{\mathrm{r}}^T.
\end{equation}
Here $\tilde {\bm{q}}_\text{rob}$ is the last row of $Q_\text{dir} R_\text{dir} R_\text{rob}^{-1} Q_\text{rob}^T$, that is $(\tilde{q}_\text{rob})_i = (Q_\text{dir} R_\text{dir} R_\text{rob}^{-1} Q_\text{rob}^T)_{d-1, i}$.

Although more convenient than the method first proposed, this latter algorithm has the limitation that the matrices $Q_\text{rob}$ and $R_\text{rob}$ depend on the specific value of $\lambda$ in the Robin boundary condition, requiring a different high precision $QR$ factorization for each value of $\lambda$. This would make this approach prohibitive for the problem considered in \cref{sec:boundary:vacuum-theoretical}, where a different value of the Robin coupling parameter $\lambda$ is satisfied by each Fourier mode, and hence $\mathcal{O}(N_x N_y/4)$ different operators are required. Changing the spatial resolution or the domain size would also forbid reusing operators, as the change in grid spacing also results in a modification in the value of $\lambda$. This can be seen from the chain rule relation
\begin{equation}
	\dod{f(x)}{x} + \lambda f(x) = g \quad \Longleftrightarrow \quad \dod{f(x')}{x'} + \underset{\lambda'}{\underbrace{\dod{x}{x'} \lambda}} f(x') = \underset{g'}{\underbrace{\dod{x}{x'} g}}.
\end{equation}
Thus, using a different grid spacing requires not only scaling of the boundary value $g$, as it was in the case of \cref{sec:fc-second-normal}, but also of the Robin coupling coefficient.

However, the aforementioned limitation can be circumvented by proposing the decomposition ${P_\text{rob} = P_\text{neu} + \lambda \hat{P}_\text{dir}}$, with $P_\text{neu}$ the matrix $P_\text{rob}$ for the special case $\lambda=0$ (corresponding to a Neumann boundary condition on the first derivative), and $\hat{P}_\text{dir}$ the Vandermonde matrix $P_\text{dir}$ but replaced with zeros in all but its last row. Substituting this decomposition in \cref{eq:P-c,eq:Prob-c} and solving for $\bm c$ leads to the equivalent relation
\begin{equation}
	(Q_\text{neu} R_\text{neu} + \lambda \hat{P}_\text{dir}) R^{-1}_\text{dir} Q^T_\text{dir} \bm{\mathrm f} = \bm{\mathrm f}_\text{rob},
\end{equation}
where, naturally, $Q_\text{neu} R_\text{neu}$ is the $QR$ factorization of $P_\text{neu}$. Noting that $\hat{P}_\text{dir}$ can be also represented as $\hat Q_\text{dir} R_\text{dir}$ with $\hat Q$ the $QR$ factorization of $P_\text{dir}$ but with zeros in all but its last row, this last expression is reduced to
\begin{equation}
	\label{fc-gram:eq:rob-before-sherman-morrison}
	(Q_\text{neu} R_\text{neu}R_\text{dir}^{-1} Q_\text{dir}^T + \lambda \hat I) \bm{\mathrm f} = \bm{\mathrm f}_\text{rob},
\end{equation}
where $\hat I$ is a matrix whose only non-zero element is $\hat I_{d-1,d-1} = 1$. This relation can then be inverted by considering $a \hat I = u^T v$ with $u = (0, \hdots, 0, \lambda)$ and $v = (0, \hdots, 0, 1)$ and using the Sherman-Morrison formula \cite{Duncan1944,Sherman1950}
\begin{equation}
	\left(A + \bm{\mathrm u} \bm{\mathrm v}^T \right)^{-1} = A^{-1} - \frac{ A^{-1}\bm{\mathrm u} \bm{\mathrm v}^T A^{-1} }{1 + \bm{\mathrm v}^T A^-1 \bm{\mathrm u} },
\end{equation}
which when applied to \cref{fc-gram:eq:rob-before-sherman-morrison} and after defining $(\tilde{\bm{q}}_\text{der})_i = (Q_\text{dir} R_\text{dir} R_\text{neu}^{-1} Q_\text{neu}^T)_{d-1, i}$, leads to
\begin{equation}
	f_{d-1} = \left(  1 - \frac{a \tilde{q}_{d-1}}{1 + \lambda \tilde{q}_{d-1}} \right) \tilde{\bm{q}}_\text{der} \bm{\mathrm f}_\text{rob}^T.
\end{equation}
The obtained value $f_{d-1}$ is a $d$-th order approximation to $f({x_{d-1})}$, a fact that can be rigorously demonstrated using the same arguments provided in \cref{sec:fc-second-normal}.

This allows efficient computation of FC-Gram transforms with Robin boundary conditions with a few pre-computed coefficients, and without extra costs when domain sizes or spatial resolutions are changed. The methods for a non-periodic function are illustrated in \cref{fc-gram:fig:fc}, where 5\textsuperscript{th} order ($d=5$) FC-Gram continuations are applied to the function $J_0(25z)e^{3z}-1$ over a grid of $N=113$ points spawning the $[0, 1]$ domain.
\begin{algorithm*}
	\DontPrintSemicolon
	\caption{Schematization of the solver for the perfectly conducting walls scenario.}\label{solver:alg:conducting}
	Starting from the field values at time $t$, ($\bm a^t$, $\bm u^t$, $p^t$, $\phi^t$), do: \;
	\nlset{1:}  \Indp \Indp Obtain intermediate variables $\bm a^{* t+\Delta t}$, $\bm u^{* t+\Delta t}$ from the pressureless momentum \; \Indp equation and the induction equation without the scalar potential.\; \Indm \Indm \Indm 
	\nlset{2:} \Indp \Indp Apply conditions to the intermediate fields: \; \Indm \Indm 
	\nlset{a.} \Indp \Indp \Indp \Indp homogeneous boundary conditions for $\bm a^{* t+\Delta t}_\parallel$; \; \Indm \Indm \Indm \Indm
	\nlset{b.} \Indp \Indp \Indp \Indp set the mean value of $a^{* t+\Delta t}_\perp$ to zero; \; \Indm \Indm \Indm \Indm
	\nlset{c.} \Indp \Indp \Indp \Indp non-slip compatible boundary conditions for the velocity. \; \Indm \Indm \Indm \Indm
	\nlset{3:} \Indp \Indp Solve two Poisson equations to remove the non-solenoidal components in $\bm u^{* t+\Delta t}$, $\bm a^{* t+\Delta t}$: \; \Indm \Indm
	\nlset{a.} \Indp \Indp \Indp \Indp for the scalar potential $\phi^{t+\Delta t}$ with homogeneous Dirichlet boundary conditions; \; \Indm \Indm \Indm \Indm
	\nlset{b.} \Indp \Indp \Indp \Indp for the pressure $p^{t+\Delta t}$ with Neumann boundary conditions in order \\ \Indp to cancel the normal velocity at the boundary at projection time. \; \Indm \Indm \Indm \Indm \Indm
	\nlset{4:} \Indp \Indp Perform a solenoidal projection to obtain the fields at the next timestep; that is:\; \Indm \Indm
	\nlset{a.} \Indp \Indp \Indp \Indp subtract $\bm \nabla \phi^{t+\Delta t}$ to $\bm a^{* t+\Delta t}$; \; \Indm \Indm \Indm \Indm
	\nlset{b.} \Indp \Indp \Indp \Indp subtract $\bm \nabla p^{t+\Delta t}$ to $\bm u^{* t+\Delta t}$. \;  \Indm \Indm \Indm \Indm
	\nlset{5:} \Indp \Indp Apply homogeneous boundary conditions to $\partial^2_{zz} \bm a^{t+\Delta t}_\parallel$ and $\partial_z a^{t+\Delta t}_\perp$. \; \Indm \Indm
	End of iteration
\end{algorithm*}

\section{A new numerical method for the incompressible MHD equations with boundaries}
\label{sec:solver}
We now present a pseudo-spectral method for evolving the incompressible MHD equations
\begin{align}
	\label{solver:eq:navier-stokes2}
	\dpd{\bm u}{t} + (\bm u \cdot \bm \nabla) \bm u &= - \nabla p - \nabla^2 \bm a \times (\bm \nabla \times \bm a) + \nu \nabla^2 \bm u, \\
	\label{solver:eq:poisson-p}
	\nabla^2 p    &= - \bm{\nabla} \cdot \left[ \left(\bm{u} \cdot \bm{\nabla}\right) {\bm u}\right], \\
	\label{solver:eq:induction-a}
	\dpd{\bm a}{t} &= \bm u \times \left( \bm \nabla \times \bm a \right) + \eta \nabla^2 \bm a - \bm \nabla \phi,\\
	\label{solver:eq:poisson-phi}
	\nabla^2 \phi &=   \bm{\nabla} \cdot \left[ \bm u \times \left(\bm \nabla \times \bm a \right)
	\right],
\end{align}
with the boundary conditions previously discussed. It should be noted that a Fourier representation is specially convenient for solving \cref{solver:eq:poisson-p,solver:eq:poisson-phi}, as inverting the Laplacian in the Fourier domain is attained by simply inverting a diagonal matrix, contrary to the denser representations found in other schemes, such as finite differences or  other spectral methods such as in classic Chebyshev polynomial decompositions \cite{Julien2009}.

In all cases, boundary conditions for the velocity field are periodic in $x$ and $y$, and no-normal velocity, no-slip in $z$ (i.e., $\bm u$ = 0 in impermeable boundaries).

\subsection{Perfectly conducting boundary conditions}
\label{sec:solver:conducting}
We consider the case with two periodic coordinates ($x$ and $y$), and a perfect conductor at $z=0$ and $L_z$. The perfect conductor boundary conditions for the vector potential were discussed in \cref{boundary:eq:bc-perfect-a,boundary:eq:bc-perfect-da,boundary:eq:bc-perfect-div,boundary:eq:bc-perfect-phi}, and are repeated here for convenience:
\begin{align}
	\label{solver:eq:apara}
	\left. \bm a_\parallel \right|_{z=0,L_z}&= \bm 0, \\
	\label{solver:eq:d2_apara}
	\left. \dpd[2]{\bm a_\parallel}{z} \right|_{z=0,L_z}&= \bm 0, \\
	\label{solver:eq:d_aperp}
	\left. \dpd{a_z}{z} \right|_{z=0,L_z}&= 0.
\end{align}

As it was explained in \cref{sec:boundary}, \cref{solver:eq:apara,solver:eq:d2_apara} are required to represent the physical boundary condition $\bm j_\parallel|_{z=0,L_z} = \bm 0$, which reduces to $\nabla^2 \bm a_\parallel|_{z=0,L_z} = \bm 0$ when the gauge condition $\bm \nabla \cdot \bm a = 0$ is fixed. Equation (\ref{solver:eq:d_aperp}) is just a gauge choice whose role is to help in maintaining the error in the gauge at the boundary small. Moreover, it should be noted that an error in the condition $(\partial_{zz}^2 \bm a_\parallel)|_{z=0,L_z}$ leads to a non-zero value for $\bm a_\parallel|_{z=0,L_z}$, as it can be seen for the evolution equation for $\bm a$ at the boundaries,
\begin{equation}
	\left. \dpd{\bm a_\parallel}{t} \right |_{z=0,L_z} = \eta \bigg [ \nabla^2 \bm a_\parallel \bigg]_{z=0,L_z}.
\end{equation}
The latter being a difussion equation, it is hence the case that any numerical error that might arise in the imposition of \cref{solver:eq:d2_apara} remains bounded and controlled when the MHD equations are evolved in time.

To impose \cref{solver:eq:apara,solver:eq:d2_apara,solver:eq:d_aperp} we now propose an explicit time-splitting integration method that imposes boundary conditions on both the tangential components of $\bm a_\parallel$ and its second normal derivative, while also utilizing a time-split scheme for the velocity field as described in \cite{Fontana2020}, based on a method presented in \cite{Kim1985,Orszag1986}. For simplicity, it will be described for a first order forward Euler time stepping, but it can be generalized to any order Runge-Kutta methods in a way fully analogous to that previously described in \cite{Fontana2020} (examples in Section \ref{sec:results} were integrated with such a second order time-splitting Runge-Kutta method).

The details of the method are schematized in \Cref{solver:alg:conducting}. In particular, starting from the wavenumber representation of the fields (obtained via FC-Gram to circumvent Gibbs phenomenon, and indicated with a hat), the pressureless momentum equation and the induction equation without the scalar potential can easily be evolved in time as
\begin{align}
	\hat{\bm u}_{nml}^{*t+\Delta t} & = \hat{\bm u}_{nml}^t + \Delta t \left[ \left[ \reallywidehat{\left(\bm u^t \cdot \bm \nabla \right) \bm u^t} \right]_{nml} - \nu k_{nml}^2 \hat{\bm u} _{nml}^t \right], \raisetag{1em} \\
	\hat{\bm a}_{nml}^{*t+\Delta t} & = \hat{\bm a}_{nml}^t + \Delta t \left[ \left[ \reallywidehat{\left(\bm u^t \times (\bm \nabla \times \bm a^t)\right)} \right]_{nml} - \nu k_{nml}^2 \hat{\bm a} _{nml}^t \right],
\end{align}
where $^*$ denotes an intermediate field (note that these fields are not solenoidal), $nml$ denote the indices of the discrete wavenumbers $k^x_n$, $k^y_m$, $k^z_l$, and $k^2_{nml} = (k^x_n)^2 + (k^y_m)^2 + (k^z_l)^2$. The non-linear terms are computed efficiently with pseudospectral calculations, that is, determining derivatives in Fourier space, products in real space and then returning to 3D wavenumber domain. After estimating the non-linear terms an exponential high cut filter is applied to reduce the mode aliasing arising from the effects of circular convolution, serving a purpose analogous to that of the well known 2/3 rule in 3D periodic solvers. This filter described in greater detail in \cite{Amlani2016,Fontana2020} rejects frequencies that would generate an aliasing error greater than the one associated to the time stepping while conserving enough high frequencies so that the periodic extension is still accurately represented, without degradation in the order of accuracy \cite{Albin2011}.

As a second step, a 1D inverse FFT is performed to recover the fields in their $(k^x,k^y,z)$ representation, after which the boundary values 
\begin{equation}
	\left. \bm a^{*t+\Delta t}_\parallel \right|_{z=0,L_z} = \bm 0,
\end{equation}
can be set by strong imposition, that is, simply setting the appropriate array values at the boundary. Regarding the velocity field, the boundary condition 
\begin{equation}
	\left. \bm u^{*t+\Delta t}_\parallel \right |_{z=0,L_z} = \Delta t \bm \nabla_\parallel p,
\end{equation} 
is enforced, which leads to an $\mathcal O(\Delta t)$ (or, more generally, $\mathcal O(\Delta t^o)$ for an $o$-th order time integration) error in the slip velocity after the projection step (see \cite{Kim1985,Orszag1986,Fontana2020}). As will be seen in Sections \ref{sec:results} and \ref{sec:convergence}, this leads to good slip velocity accuracy for the time integration schemes and timesteps used in typical turbulent MHD simulations. 

Following the imposition of the boundary conditions for the intermediate fields, the variables are transformed back to their $(k^x,k^y,k^z)$ representation, after which the condition ${a_z^{*t+\Delta t}(k^x=0, k^y=0, k^z=0)=0}$ is also enforced for improved stability in very long time integrations. Note that this corresponds to fixing a constant in $a_z$ (i.e., to a gauge freedom) and has no effect on the magnetic field. 

Then, at the third step, Poisson equations
\begin{align}
	\nabla^2 p^{t+\Delta t} &= \frac{1}{\Delta t} \bm \nabla \cdot \bm u^{*t+\Delta t},\\
	\nabla^2 \phi^{t+\Delta t} &= \frac{1}{\Delta t} \bm \nabla \cdot \bm a^{*t+\Delta t},
\end{align}
for the pressure and electric scalar potential are solved. The inhomogeneous solutions for both $\phi$ and $p$, $\phi^I$ and $p^I$, respectively, can easily be obtained in the 3D wavenumber domain as
\begin{equation}
	\label{solver:eq:inhomogeneous}
	\begin{Bmatrix}
		\phi^{I t+\Delta t}_{nml} \\[.8em]
		p^{I t+\Delta t}_{nml}
	\end{Bmatrix} = - \frac{i \bm k_{nml}}{k_{nml}^2} \cdot \begin{Bmatrix}
	\hat{\bm a}^{*t+\Delta t}_{nml}\\[.8em]
	\hat{\bm u}^{*t+\Delta t}_{nml}
\end{Bmatrix}
\end{equation}
which are, naturally, defined up to constants $\phi^{I}_{0,0,0}$, $p^{I}_{0,0,0}$. On the other hand, in this geometry, the homogeneous solutions $\phi^H$, $p^H$ can be easily constructed from an explicit representation in the $(k_x, k_y, z)$ domain, that is
\begin{multline}
	\label{solver:eq:laplace}
	\begin{Bmatrix} \hat \phi^{H t+\Delta t}_{nm} (z) \\[.8em]
	\hat p^{H t+\Delta t}_{nm} (z)
	\end{Bmatrix} = 
	\begin{Bmatrix} A_{nm} \\[.8em] 
	A'_{nm}
	\end{Bmatrix}e^{\gamma_{nm}(z-L_z)} 
	+ \begin{Bmatrix} B_{nm} \\[.8em] 
	B'_{nm} 
	\end{Bmatrix}	e^{-\gamma_{nm}z} \\
	+ \begin{Bmatrix} C + Dz \\[.8em] 
	C' + D'z
	\end{Bmatrix}.
\end{multline}
The non-primed coefficients in the former expression are derived from the boundary conditions $\phi|_{z={0,L_z}} = 0$ as
\begin{align}
    A_{mn} &= \frac{\phi^{It+\Delta t}_{nm}(z=0)e^{-\gamma_{nm}L_z} - \phi^{It+\Delta t}_{nm}(z=L_z)}{1-e^{-2\gamma_{nm}L_z}},\\
    B_{mn} &= \frac{\phi^{It+\Delta t}_{nm}(z=L_z)e^{-\gamma_{nm}L_z} - \phi^{It+\Delta t}_{nm}(z=0)}{1-e^{-2\gamma_{nm}L_z}},\\
    C      &= - \phi^{It+\Delta t}_{0,0}(z=0),\\
    D      &= \frac{1}{L_z} \left[ \phi^{It+\Delta t}_{0,0}(z=0) - \phi^{It+\Delta t}_{0,0}(z=L_z) \right],
\end{align}
whereas $(\partial_z p)|_{z=0,L_z} = u_z^{*t+\Delta t}/\Delta t$ (to cancel the wall normal velocity) leads to the similar relations
\begin{align}
	\label{solver:eq:pressure-A}
	A'_{mn} &= t'_{nm}-b'_{nm}e^{-\gamma_{nm}L_z},\\
   	\label{solver:eq:pressure-B}
	B'_{mn} &= t'_{nm}e^{-\gamma_{nm}L_z} - b'_{mn},\\
   	\label{solver:eq:pressure-C}
	C'      &= 0,\\
   	\label{solver:eq:pressure-D}
	D'      &= \frac{1}{\Delta t} b'_{00} = \frac{1}{\Delta t} t'_{00},
\end{align}
with
\begin{align}
	\label{solver:eq:pressure-b}
	b'_{nm} &= \hat{\bm z} \cdot \left[ \frac{\hat {\bm u}^{*t+\Delta t}_{nm} (z=0)}{\Delta t}- \bm \nabla \hat{p}^{It+\Delta t}_{nm}(z=0) \right] \Gamma, \\
	\label{solver:eq:pressure-t}
	t'_{nm} &=	\hat{\bm z} \cdot \left[ \frac{\hat{\bm u}^{*t+\Delta t}_{nm} (z=L_z)}{\Delta t}- \bm \nabla \hat{p}^{It+\Delta t}_{nm}(z=L_z) \right] \Gamma,\\
	\Gamma &= \frac{1}{\gamma_{nm}(1-e^{-2\gamma_{nm}L_z})}.
\end{align}

\begin{algorithm*}
	\DontPrintSemicolon
	\caption{Schematization of the solver for a magnetofluid surrounded by vacuum.}\label{solver:alg:vacuum}
	Starting from the field values at time $t$, ($\bm a^t$, $\bm u^t$, $p^t$, $\phi^t$), do: \;
	\nlset{1:}  \Indp \Indp Obtain intermediate variables $\bm a^{* t+\Delta t}$, $\bm u^{* t+\Delta t}$ from the pressureless momentum \; \Indp equation and the induction equation without the scalar potential.\; \Indm \Indm \Indm 
	\nlset{2:} \Indp \Indp Apply conditions to the intermediate fields: \; \Indm \Indm 
	\nlset{a.} \Indp \Indp \Indp \Indp set the mean value of $a^{* t+\Delta t}_\perp$ to zero; \; \Indm \Indm \Indm \Indm
	\nlset{b.} \Indp \Indp \Indp \Indp non-slip compatible boundary conditions for the velocity. \; \Indm \Indm \Indm \Indm
	\nlset{3:} \Indp \Indp Solve two Poisson equations to remove the non-solenoidal components in $\bm u^{* t+\Delta t}$, $\bm a^{* t+\Delta t}$: \; \Indm \Indm
	\nlset{a.} \Indp \Indp \Indp \Indp for the scalar potential $\phi^{t+\Delta t}$ with Robin boundary conditions; \; \Indm \Indm \Indm \Indm
	\nlset{b.} \Indp \Indp \Indp \Indp for the pressure $p^{t+\Delta t}$ with Neumann boundary conditions in order \\ \Indp to cancel the normal velocity at the boundary at projection time. \; \Indm \Indm \Indm \Indm \Indm
	\nlset{4:} \Indp \Indp Perform a solenoidal projection to obtain the fields at the next timestep; that is:\; \Indm \Indm
	\nlset{a.} \Indp \Indp \Indp \Indp subtract $\bm \nabla \phi^{t+\Delta t}$ to $\bm a^{* t+\Delta t}$; \; \Indm \Indm \Indm \Indm
	\nlset{b.} \Indp \Indp \Indp \Indp subtract $\bm \nabla p^{t+\Delta t}$ to $\bm u^{* t+\Delta t}$. \;  \Indm \Indm \Indm \Indm
	\nlset{5:} \Indp \Indp Apply Robin boundary conditions to $\bm a^{t+\Delta t}$. \; \Indm \Indm
	End of iteration
\end{algorithm*}

The homogeneous solutions to the scalar potential and the pressure, along with their derivatives along $z$ (whose exact expressions are easily obtained), can be FC-Gram transformed to the $k^x,k^y,k^z$ domain, obtaining the full pressure and electrostatic potential as $\hat p_{nml} = \hat p^I_{nml} + \hat p^H_{nml}$ and $\hat \phi_{nml} = \hat \phi^I_{nml} + \hat \phi^H_{nml}$ respectively. The intermediate fields can be therefore projected at the fourth step onto the solenoidal space simply as
\begin{align}
	\label{solver:eq:projection-a}
	\hat{\bm a}^{t+\Delta t}_{nml} = \hat{\bm a}^{*t+\Delta t}_{nml} - \Delta t \nabla \hat \phi^{t+\Delta t}_{nml}, \\
	\label{solver:eq:projection-v}
	\hat{\bm u}^{t+\Delta t}_{nml} = \hat{\bm u}^{*t+\Delta t}_{nml} - \Delta t \nabla \hat p^{t+\Delta t}_{nml}.
\end{align}

At this point the velocity field at the next time step is fully determined. To complete the calculation of $\bm a^{t+\Delta t}$, the fifth and final step is carried out, so that its boundary values must be still adjusted, namely such that $(\partial^2_{zz} \bm a_\parallel)|_{z=0,Lz} = 0$ and $(\partial_{z} a_z)|_{z=0,Lz} = 0$. This can be easily accomplished by returning to the $(k_x,k_y,z)$ domain and applying the method introduced in \cref{sec:fc-second-normal} to obtain the appropriate boundary values for $\bm a_\parallel$, whereas the related Neumann projector described in \cref{sec:fc-robin} (Robin boundary conditions with $\lambda=0$) is employed to adjust $a_z$ at the endpoints. In all the cases the computed values are set using strong imposition. A final FC-Gram transformation of $\bm a$ in the $\hat{\bm z}$ direction leaves the fields ready for the calculations of the next time step. 

\subsection{Vacuum boundary conditions}
A very similar approach to the one introduced in the previous subsection can be employed for the case of a magnetofluid periodic in $x$ and $y$ and surrounded by vacuum on two sides in $z$. The electromagnetic and velocity boundary conditions are repeated here for practicality
\begin{align}
	\label{solver:eq:robin_a}
	\left. \left( \dod{\bm a^{\I}_{nm}}{z} + \gamma_{nm} \bm a^{\I} \right ) \right|_{z=0,L_z} = \bm 0,\\
	\label{solver:eq:robin_phi}
	\left. \left( \dod{\phi^{\I}_{nm}}{z} + \gamma_{nm} \phi^{\I} \right ) \right|_{z=0,L_z} = 0,\\
	\label{solver:eq:robin_velocity}
	\left. \bm u \right|_{z=0,L_z} = \bm 0.
\end{align}
Contrary to the conducting case, there is no consistency condition to impose, and the simplest possible approach is to directly enforce  \cref{solver:eq:robin_a,solver:eq:robin_phi} at the end of each time step, whereas \cref{solver:eq:robin_velocity} is enforced, as before, at the intermediate step. As a result, and as it can be easily noted in \Cref{solver:alg:vacuum}, the scheme is essentially the same as that discussed in the previous subsection, with the exception that there is no need to Fourier transform the intermediate magnetic potential. The only condition imposed in $a^{*t+\Delta t}$ is $a_z^{*t+\Delta t}(k^x=0, k^y=0, k^z=0) = 0$, for the same considerations mentioned in the previous section. As will be shown in \cref{sec:convergence}, the procedure works well and errors are seen to decrease rapidly as spatial resolution is increased, and as the time step is decreased. 

Solving the Poisson equation for $p^{t+\Delta t}$ is an identical process to the one described in \cref{solver:eq:inhomogeneous,solver:eq:laplace,solver:eq:pressure-A,solver:eq:pressure-B,solver:eq:pressure-C,solver:eq:pressure-D,solver:eq:pressure-b,solver:eq:pressure-t}. Similarly, the inhomogeneous electric scalar potential is again obtained from \cref{solver:eq:inhomogeneous} while the homogeneous contribution satisfying homogeneous Robin boundary conditions can be obtained from the closed form solution
\begin{equation}
	\label{solver:eq:robin_laplace}
	\hat \phi^{H t+\Delta t}_{nm} (z) = A_{nm} e^{\gamma_{nm}(z-L_z)} + B_{nm} e^{-\gamma_{nm}z} + C + Dz,
\end{equation}
where
\begin{align}
	A_{nm} &= -\frac{1}{2\gamma_{nm}} \left(\partial_z \phi^{I t+\Delta t}_{nm} (z=L_z) + \gamma_{nm} \phi^{I t+\Delta t}_{nm}(z=L_z) \right),\\
	B_{nm} &= \frac{1}{2\gamma_{nm}} \left(\partial_z \phi^{I t+\Delta t}_{nm}(z=0) - \gamma_{nm} \phi^{I t+\Delta t}_{nm}(z=0) \right),\\
	C      &= 0,\\
	D      &= \partial_z \phi^{I t+\Delta t}_{0,0}(z=0).
\end{align}
After adding the homogeneous and inhomogeneous solutions, the projections onto the respective solenoidal spaces are carried out utilizing \cref{solver:eq:projection-v,solver:eq:projection-a}.

Finally, similarly to the perfectly conducting case, the boundary conditions in the fifth step are imposed by obtaining the $(k^x,k^y,z)$ representation of the magnetic vector potential, but then instead using the method introduced in \cref{sec:fc-robin} to efficiently and accurately obtain the appropriate boundary values for the Robin case. Finally $\bm a^{t+\Delta t}$ is transformed back to the $k^x, k^y, k^z$ domain with a FC-Gram operation, at which point the computation of variables at the next time step can begin.
\section{Application to a Hartmann flow scenario}
\label{sec:results}
As an application we consider the study of a Hartmann flow \cite{Davidson2001,Moreau2007}, which will also serve as a validation case of the numerical method, as well as the accuracy of the boundary conditions (see more details in Section \ref{sec:convergence}). For that purpose the solvers presented in \cref{sec:solver} were implemented atop the FC-Gram based PDE solver \texttt{SPECTER} (freely available at \url{https://www.github.com/mfontanaar/SPECTER}), which uses a hybrid MPI-OpenMP-CUDA parallelization for efficiently running in computer clusters \cite{Mininni2011,Rosenberg2020}. Simulations were carried out inside a dimensionless $xy$-periodic box of size $L_x\times L_y \times L_z = 2\pi \times \pi \times 1$ and for varying resolutions and parameters. In the $\hat{\bm z}$ direction both conducting and vacuum boundaries were explored.

\begin{figure}
	\centering
	\includegraphics[width=\linewidth, keepaspectratio]{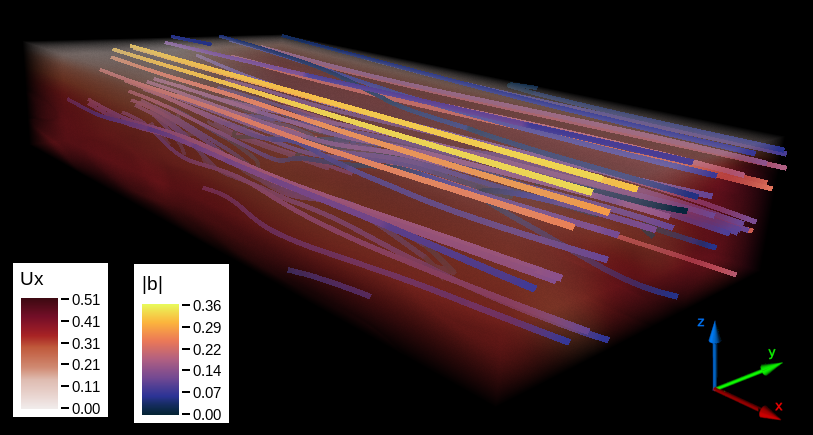}
	\caption{Rendering of the region $(0,0,1/2)\times(\pi, \pi/2, 1)$ for simulation C24 at $t=162$. The volume rendering displays the streamwise velocity $u_x$ and also shown are field lines for $\bm b$ colored using $|\bm b|$ to denote the local magnetic field intensity. Note that the induced magnetic field in this case is confined inside the domain.}
	\label{results:fig:conducting}
\end{figure}

\subsection{Theoretical and experimental background}
\label{subsec:theoretical}
The Hartmann problem consists of a flow driven by a mean pressure gradient $G$ between the open ends of a channel, and with a stationary uniform magnetic field $\bm B_0$ applied perpendicular to the channel walls. In particular, we consider the mean pressure gradient to be in the $-\hat{\bm x}$ direction, whereas $\bm B_0 = b_0 \hat{\bm z}$. The total magnetic field is then given by $\bm B = \bm B_0 + \bm b$, with the dynamic magnetic field $\bm b = \bm \nabla \times \bm a$ satisfying the boundary conditions described before depending on the wall properties at $z=0$ and $z=L_z$. This problem was first experimentally studied by Hartmann and Lazarus \cite{Hartmann1937a,Hartmann1937b}, and is a setting that remains relevant to the present day in industrial applications as well as a benchmark for numerical methods in MHD \cite{Shercliff1953,Cuevas1997,Krasnov2004,Sarris2007,Vantieghem2009,Zikanov2014,Krasnov2012,Llorente2019}.

The evolution equations for $\bm u$ and $\bm a$ are given by
\begin{align}
	\label{results:eq:mhd-momentum}
	\dpd{\bm u}{t} &= \begin{aligned}[t]- \bm \nabla p -  (\bm{u} \cdot \bm \nabla ) \bm u - \nabla^2 \bm a \times \left[ (\bm \nabla \times \bm a) + \bm B_0 \right] \\
	+ \nu \nabla^2 \bm u + G \hat{\bm x},
	\end{aligned}\\
	\label{results:eq:mhd-induction}
	\dpd{\bm a}{t} &= - \bm \nabla \phi + \bm u \times \left[(\bm \nabla \times \bm a) + \bm B_0 \right]+ \eta \nabla^2 \bm a.
\end{align}
It is convenient to characterize the system in terms of the dimensionless Hartmann number $Ha$, centerline Reynolds number $Re_0$, and the magnetic Prandtl number $Pm$, which are defined respectively as
\begin{equation}
	Ha = \frac{b_0 \delta}{\sqrt{\eta \nu} }, \qquad \qquad Re_0 = \frac{u_0 \delta}{\nu}, \qquad \qquad Pm = \frac{\nu}{\eta},
\end{equation}
where $u_0$ is the characteristic flow speed at the center of the channel, and $\delta = L_z/2$ is the box half-height. Physically, $Ha$ and $Re_0$ represent the inverse of weighing the viscous effects to those produced by the Lorentz force and inertia, respectively, whereas $Pm$ is an estimate of the ratio between mechanical and ohmic diffusion timescales. Additionally, a magnetic Reynolds number $Rm$ can be defined as $Rm = Re_0/Pm$, which is a ratio between the advection of the magnetic field against the ohmic diffusion.

Numerically speaking, the most challenging regime to explore is the one where both the velocity and induced magnetic fields contain significant energy across a large range of scales, i.e., when both fields are turbulent. This regime is attained for $Re_0 \gg 1$ and $Rm \gg 1$. It should be noted that, traditionally, the Hartmann channel is numerically studied in the $Rm \ll 1$ limit \cite{Cuevas1997,Krasnov2012}. For this regime, a quasistatic approximation to the MHD equations is employed, which reduces the problem stiffness. However, one advantage of our method is that it will allow us to evolve in time the turbulent behavior of the magnetic field. Consequently, to test the method in this challenging regime without requiring prohibitive computational resources, the case $Pm=1$ will be considered henceforth.

For the case of a non-conducting fluid ($\bm a = \bm B_0 = 0$), that is, a plane Poiseuille flow, the dynamics encompassed in Eq.~(\ref{results:eq:mhd-momentum}) are well understood. A laminar parabolic profile is found up to a critical Reynolds number $Re_c$. For $Re_0 > Re_c$ a turbulent flow develops except in a mechanical boundary layer, whose thickness shrinks as $Re_0$ increases\cite{Kim1987,Lee2015}. For a conducting fluid, however, the externally imposed magnetic field induces currents in the spanwise direction, which in turn increase drag in the fluid via ohmic dissipation, a phenomenon sometimes called \textit{magnetic braking} \cite{Davidson2001,Moreau2007}. Then, if the magnetic field is strong, or more precisely, if $Ha \gg 1$, the flow develops a core motion (which might or might not be turbulent in nature) and a magnetic boundary layer forms. This last region is commonly known as the Hartmann layer, and its thickness $\delta_{Ha}$ scales as $Ha^{-1}$.

Also when $Ha \gg 1$, and contrary to the hydrodynamic case, the presence of turbulence in the bulk of the flow is not directly controlled by the classical Reynolds number nor by the interaction number $N=Ha^2/Re$, but by the modified Reynolds number $R$ defined as
\begin{equation}
	R = \frac{Re_0}{Ha}.
\end{equation}
It should be noted that $R$ is none other than the Reynolds number at the Hartmann scale. For the case with $Pm \ll 1$ (i.e., $Rm \ll 1$), linear stability analysis suggests a subcritical transition to turbulence at $Re_c \approx 5\times10^4$ \cite{Lock1955,Moreau2007}. However, experiments and simulations at $Ha\gg1$ and $R_m\ll1$ place this limit in the $200 < Re_c < 400$ region \cite{Moresco2004,Krasnov2012}, and thus finite amplitude perturbations must be considered when analyzing stability.

In the laminar case with $R_m \ll 1$, a closed-form solution to the Hartmann flow is given by the streamwise velocity $\bm u = u_x(z) \hat{\bm x}$, with
\begin{equation}
	\label{results:eq:u}
	u_x(z) = u_0 \left[ 1 - \frac{\cosh \left( (z-\delta)/\delta_{Ha} \right)}{\cosh \left(\delta/\delta_{Ha} \right)} \right], 
\end{equation}
for a channel whose walls are at $z=0$ and at $z=2\delta$. As before, $u_0$ is the velocity at the center of channel.

\subsection{Simulations with perfectly conducting walls}
\label{sec:results:conducting}
\begin{figure}
	\centering
	\includegraphics[width=.8\linewidth, keepaspectratio]{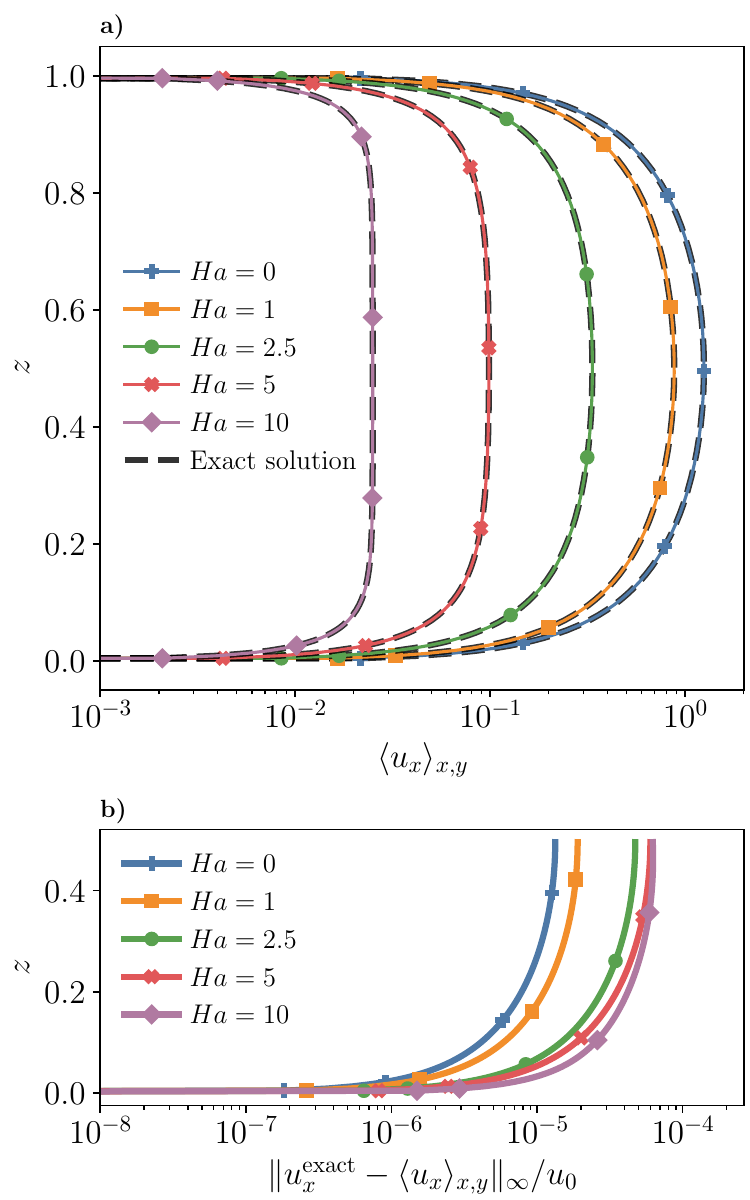}
	\caption{Results for simulations C0 to C4 after 200 eddy turnover times. \textbf{a)} Mean vertical profile for the streamwise velocity $\langle u_x \rangle_{x,y}$, alongside black dashed lines which represent the corresponding closed-form laminar solution for each value of $Ha$. \textbf{b)} Relative maximum difference between the mean profiles and the closed-form solution as a function of $z$. In both cases, profiles for varying values of $Ha$ are denoted employing different colors and markers.}
	\label{results:fig:conducting_varying_b0}
\end{figure}

\begin{table*}
	\begin{center}
		\small
	\begin{tabular}{c c c c c S[table-format=-1.2e-1] S[table-format=-1.1e-1] S[table-format=-1.1e-1]}
		\toprule
		\toprule
		Run ID	& $N_x\times N_y\times N_z$ & ${Ha}$ & ${Re_0}$  & ${R}$ &  $b_{0}$ &  ${G}$   &${\nu=\eta}$\\
		\midrule
C0  & $128 \times  64 \times  231$ &    0 &  125 & $\infty$ &    0e+0 &    5e-2 &    5e-3\\
C1  & $128 \times  64 \times  231$ &    1 &   88 &       88 &    1e-2 &    5e-2 &    5e-3\\
C2  & $128 \times  64 \times  231$ &  2.5 &   33 &       13 &  2.5e-2 &    5e-2 &    5e-3\\
C3  & $128 \times  64 \times  231$ &    5 &   10 &        2 &    5e-2 &    5e-2 &    5e-3\\
C4  & $128 \times  64 \times  231$ &   10 & 2.50 &     0.25 &    1e-1 &    5e-2 &    5e-3\\
C21 & $128 \times  64 \times  231$ &  2.5 &   67 &       27 & 1.25e-2 &  2.5e-2 &  2.5e-3\\
C22 & $128 \times  64 \times  231$ &  2.5 &  168 &       67 &    5e-3 &    1e-2 &    1e-3\\
C23 & $128 \times  64 \times  231$ &  2.5 &  916 &      366 & 1.25e-3 &  2.5e-3 &  2.5e-4\\
C24 & $256 \times 128 \times  487$ &  2.5 & 2885 &     1154 &    5e-4 &    1e-3 &    1e-4\\
C31 & $128 \times  64 \times  231$ &    5 &   20 &        4 &  2.5e-2 &  2.5e-2 &  2.5e-3\\
C32 & $128 \times  64 \times  231$ &    5 &   49 &       10 &    1e-2 &    1e-2 &    1e-3\\
C33 & $128 \times  64 \times  231$ &    5 &  200 &       40 &  2.5e-3 &  2.5e-3 &  2.5e-4\\
C34 & $128 \times  64 \times  231$ &    5 &  822 &      164 &    1e-3 &    1e-3 &    1e-4\\
C35 & $128 \times  64 \times  231$ &    5 & 2178 &      436 &    5e-4 &    5e-4 &    5e-5\\
C36 & $256 \times 128 \times  487$ &    5 & 4877 &      975 &  2.5e-4 &  2.5e-4 &  2.5e-5\\
		\bottomrule
		\bottomrule
	\end{tabular}
	\caption{Summary of the simulations performed for the perfectly conducting case. In all the cases a number of continuation points $C = 25$ was employed, as well as 9\textsuperscript{th} order FC-Gram extensions ($d = 10$). In all the simulations the fluid spawns the domain $L_x \times L_y \times L_z = 2\pi \times \pi \times 1$. ``Run ID" labels each run, $N_x\times N_y\times N_z$ gives the linear resolution, $Ha$ is the Hartman number, $Re_0$ the Reynolds number, $R$ the modified Reynolds number, $b_0$ the amplitude of the external uniform magnetic field, $G$ the external pressure gradient, and $\nu=\eta$ are the dimensionless kinematic viscosity and magnetic diffusivity.}
	\label{results:tbl:conducting}
	\end{center}
\end{table*}
We now report the simulations performed for the case in which the walls are made of a perfectly conducting material. A summary of all the simulations performed for this case can be found in \cref{results:tbl:conducting}. A graphical representation of the 3D fields, obtained using the software \texttt{VAPOR} \cite{Clyne2007,Vapor}, is shown in \cref{results:fig:conducting}. Note magnetic field lines correspond to the $\bm b$ field, i.e., to the field induced by the fluid motion, and not to the total magnetic field $\bm B_0 + \bm b$.

We first consider simulations C0 to C4, for which both the pressure gradient and the diffusivities were kept constant and only $b_0$ was varied. Even more, values for $G$ and $\nu$ were selected in order to achieve a centerline velocity $u_0$ of $\mathcal O (1)$ for the case $b_0 = 0$. Initial conditions for this set of simulations consist of random velocity and vector potential fields whose energy is concentrated at the largest scales of the system and equal to $10^{-2}$. The idea is to have a small perturbation to see if the laminar profile establishes, or if the system develops other solution instead. Additionally, these initial fields are solenoidal and satisfy the appropriate boundary conditions.

In \cref{results:fig:conducting_varying_b0} we present the results obtained for the set of simulations C0 to C4 after 200 eddy turnover times. More precisely, the mean vertical profile for the streamwise velocity $\langle u_x \rangle_{x,y}$ (where the subindices in the brackets indicate averages are performed over the $x$ and $y$ coordinates) is shown for each run, alongside the relative maximum error between the simulated profile and the closed-form solution as a function of $z$. In particular, for C0 (with $b_0 = Ha = 0$) the plane Poiseuille flow solution $u(z) = G z(L_z-z)/(2\nu)$ is recovered \cite{Pope2000,Fontana2020}. Instead, simulations C1 to C4, in all the cases, accurately fit the closed-form solution in \cref{results:eq:u}, while also satisfying the theoretical prescription for the centerline velocity
\begin{equation}
	u_0= \frac{G \nu}{b_0^2} ,
	\label{results:eq:u0-conductor}
\end{equation}
given in \cite{Davidson2001} for the case of conducting walls. It should be mentioned, however, that \cref{results:eq:u0-conductor} is derived for the limit of high magnetic diffusivity ($Rm$ and $Pm \ll 1$), and care should be taken when considering it as the exact laminar solution for moderate values of $Rm$. In spite of this, note the solution is statisfied with small errors, and that the no-slip boundary condition for the velocity in the walls is also well satisfied (more on these errors will be discussed in Section \ref{sec:convergence}). Moreover, for all these simulations, the standard deviation in the streamwise and spanwise directions is of order $10^{-8}$ at $t=200$, indicating the solutions are significantly homogeneous in the $\hat{\bm x}$ and $\hat{\bm y}$ directions, as expected in this regime.

We then proceed to select two distinct values of $Ha$, namely $Ha=2.5$ (run C2) and $Ha=5$ (run C3), in order to test the flow stability at fixed $Ha$ as $Re_0$ and $R$ increase. These values for $Ha$ were chosen because they correspond to two qualitatively different streamwise velocity profiles: whereas in C2 it still resembles a parabola ($\delta_{Ha}=2\delta/5$), run C3 already has a clear uniform core in the streamwise velocity ($\delta_{Ha}=\delta/5$). To study the flow stability, a 10\% of random noise in the largest scales is added to the last output of the velocity field of simulations C2 and C3. These perturbed velocity fields, together with the last output of $\bm a$ in each run, are used as initial conditions for simulations C21 and C31 respectively. The new set of parameters ($\nu'$, $G'$, and $b_0'$) are obtained from the previous ones ($\nu$, $G$ and, $B_0$) by keeping constant both $Ha$ and the analytical prediction for $u_0$, so that fixing $\nu'=\nu/2$ results in $G' = G/2$ and $b_0' = b_0/2$. An analogous procedure is employed to initialize simulations C22 and C32 respectively from the last states of C21 and C31, and so on for simulations C23, C24, C33, and C34.

\begin{figure}[t]
	\centering
	\includegraphics[width=.95\linewidth, keepaspectratio]{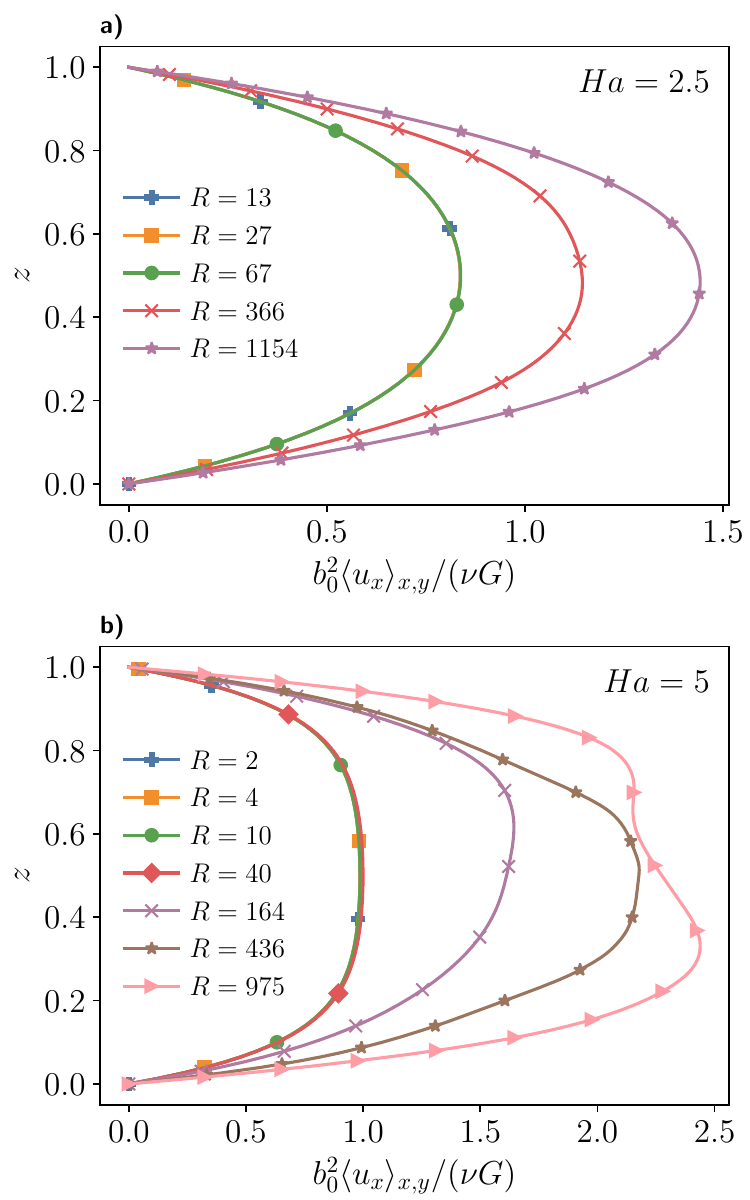}
	\caption{\textbf{a)} Mean vertical profile for the streamwise velocity normalized by the analytical prescription for $u_0$, $b_0^2 \langle u_x \rangle_{x,y}/(\nu G)$, as a function of $z$ for the set of simulations C2X ($Ha=2.5$). \textbf{b)} Same for simulations C3X ($Ha=5$). In all the cases, profiles for different values of $R$ are differentiated by their colors and markers.}
	\label{results:fig:conducting_fixed_ha}
\end{figure}

\begin{figure}[t]
	\centering
	\includegraphics[width=\linewidth, keepaspectratio]{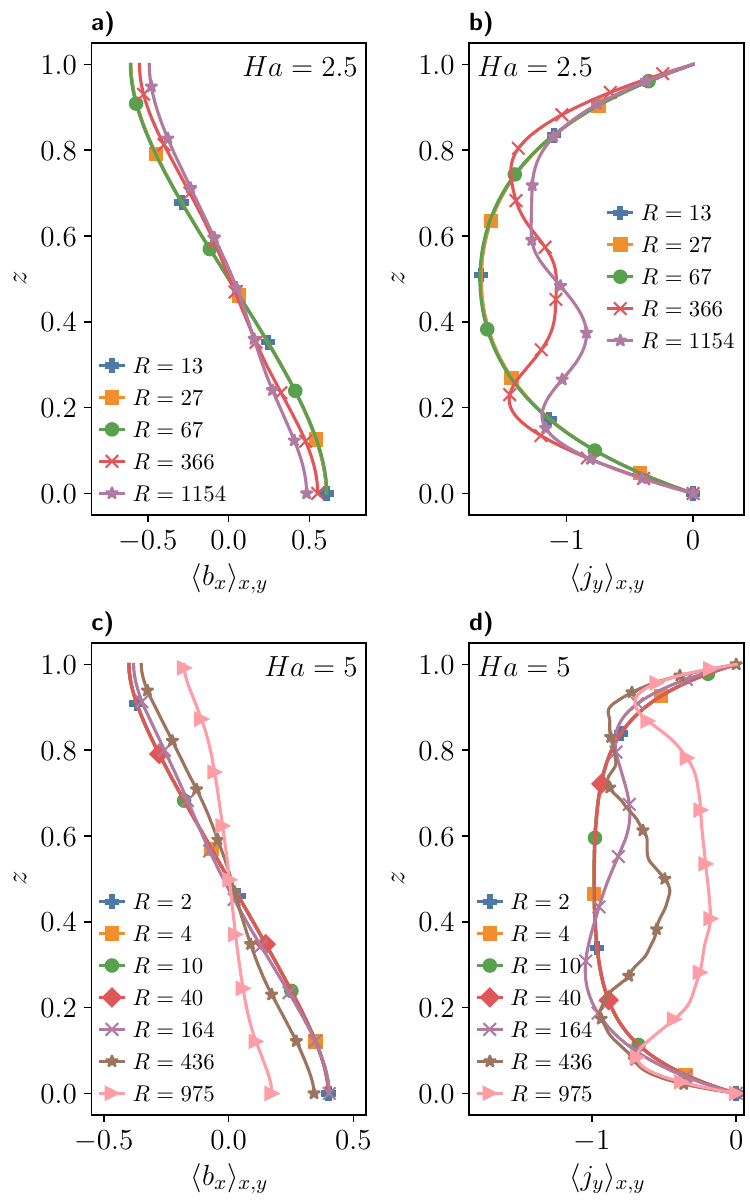}
	\caption{Left column: mean vertical profile for the streamwise induced magnetic field $\langle b_x \rangle_{x,y}$ as a function of $z$ for simulations C2X (a) and C3X (c). Right column: mean vertical profile for the spanwise current density $\langle j_y \rangle_{x,y}$ as a function of $z$ for simulations C2X (b) and C3X (d). In all cases profiles for different values of $R$ are differentiated by their colors and markers.}
	\label{results:fig:conducting_magnetic}
\end{figure}

In \cref{results:fig:conducting_fixed_ha} the mean streamwise velocity profiles at fixed $Ha$ are shown, normalized by the theoretical centerline velocity in \cref{results:eq:u0-conductor}. It can be readily observed that for the C2X runs ($Ha=2.5$) a departure from the laminar solution is found for $R_c$ in the range 67--366, which is compatible with the critical values for $R$ reported in \cite{Moresco2004,Krasnov2012} for the quasistatic regime ($Pm \ll 1$). A similar behavior is seen for the C3X simulations ($Ha=5$) where the same kind of transition is observed but for $40 < R_c < 164$. Moreover, and as expected from the theoretical discussion in \cref{subsec:theoretical}, \cref{results:fig:conducting_fixed_ha} shows that the solutions for $R > R_c$ present higher velocity values than those expected for the laminar case, as magnetic braking is no longer the dominating dissipation mechanism. For the values of $R > R_c$ explored, up to a doubling of the centerline velocity is observed when compared to the analytical solution.

For the same simulations, a change is observed with $R$ in both the streamwise magnetic field as well as the spanwise current density, as seen in \cref{results:fig:conducting_magnetic} (note also in this figure how the boundary conditions are satisfied as $j_y$ goes to zero in the boundaries; see errors in Section \ref{sec:convergence}). In this figure, the streamwise magnetic field monotonically diminishes as a function of $R$ for $R > R_c$. The decrease is more pronounced for larger $Ha$; it is of $\approx 60\%$ for $Ha=5$ and $R=872$, whereas for $Ha=2.5$ and $R=1185$ it diminishes by $\approx 33\%$ with respect to the laminar value. For the spanwise current density we observe a parabolic profile for $R < R_c$, whereas the transitional regime is characterized by a sharp drop in the current for $|z-\delta| > \delta_\text{Ha}$, that is, outside the Hartmann boundary layer. This latter phenomenon also showcases that the induced current in the core of the flow cannot produce enough magnetic braking in order for $u_x$ to obey \cref{results:eq:u}. Even more relevant for evaluating the robustness of the proposed solver, for $R > R_c$ all $u_x$, $b_x$ and $j_y$ present standard deviation values comparable to their characteristic values. 

\begin{figure}[h!]
	\centering
	\includegraphics[width=\linewidth, keepaspectratio]{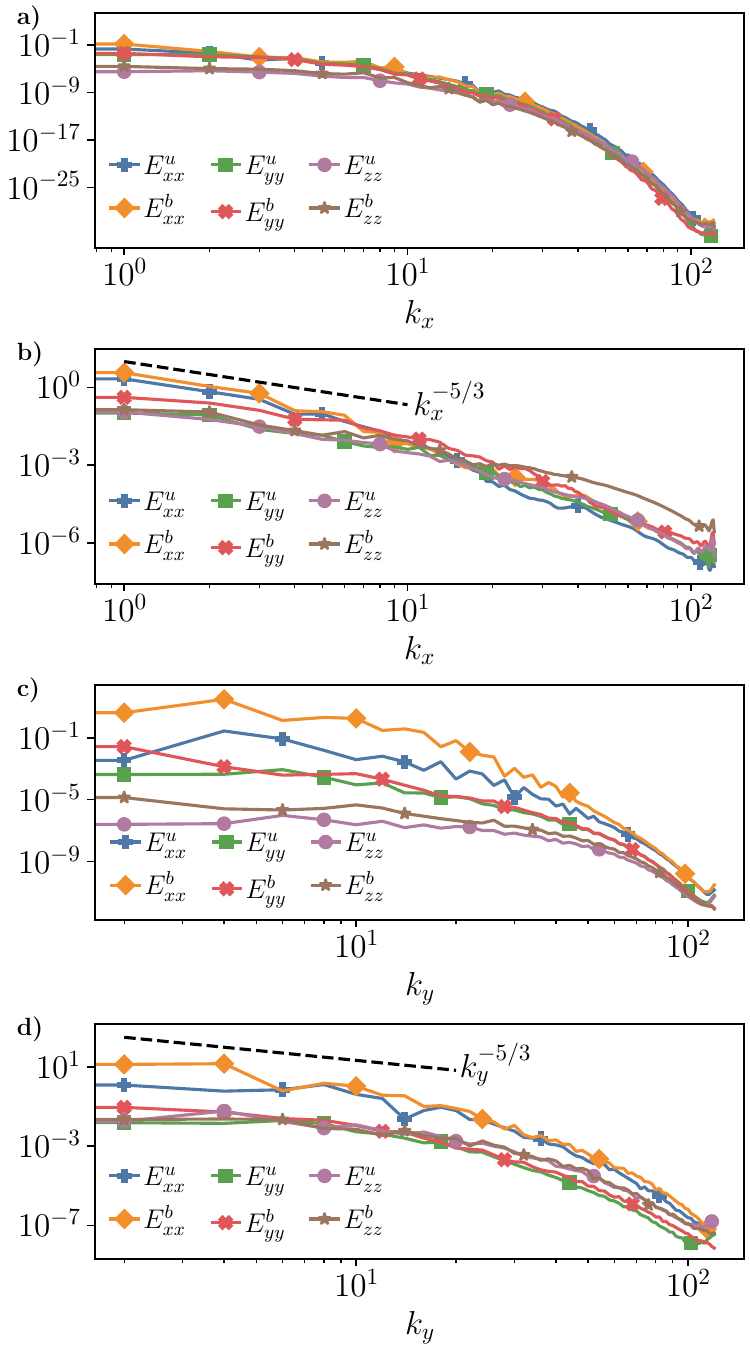}
	\caption{Kinetic and magnetic spectra for simulation C24 after 160 eddy turnover times. Panels \textbf{a)} and \textbf{b)} show the spectrum for all $x$, $y$ and $z$ components of both the velocity and magnetic field as a function of the streamwise wavenumber $k_x$. Spectra both near the wall (a) and at the center of the channel (b) are shown, computed at distances $z=0.01$ and $z=0.5$ of the bottom wall, respectively. Similarly, panels \textbf{c)} and \textbf{d)} show the spectrum for all the components of the velocity and magnetic fields as a function of the spanwise wavenumber $k_y$. Spectra both near the wall (c) and at the center of the channel (d) are shown, computed at distances $z=0.01$ and $z=0.5$ of the bottom wall, respectively. In all the panels distinct markers and line colors are used to differentiate spectra.}
	\label{results:fig:conducting_spectra}
\end{figure}

Finally, we also consider the 1D spectra for every component of the field as a function of both the streamwise as well as spanwise wavenumbers, defined for the velocity field as
\begin{align}
	E^{u}_{ii} (k_x, z) &= \sum_{k_y} \abs{\hat u_i (k_x, k_y, z)}^2,\\
	E^{u}_{ii} (k_y, z) &= \sum_{k_x} \abs{\hat u_i (k_x, k_y, z)}^2,
\end{align}
and with a similar expression for the magnetic spectra $E^b_{ii}$. The results are shown in \cref{results:fig:conducting_spectra} for run C24, at two qualitatively distinct vertical slices: one near the bottom wall ($z = 5 \Delta z = 0.01$) and the other in the center of the channel ($z=0.5$). First and foremost, it should be noted that in both cases the spectra seems to be appropriately resolved, with no noticeable aliasing. Although near the walls the spectra is sharper, especially as a function of the streamwise wavenumber, in the center of the channel inertial range regions compatible with a $k^{-5/3}$ spectrum can be observed. The method can thus capture solutions with the proper boundary conditions even in the turbulent regime with small-scale field fluctuations.

\subsection{Simulations with vacuum boundary conditions}
\begin{figure}
	\centering
	\includegraphics[width=\linewidth, keepaspectratio]{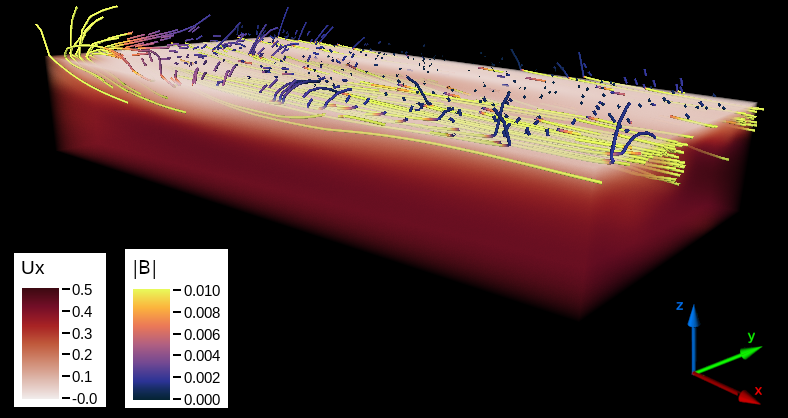}
	\caption{Rendering of the region $(0,0,0)\times(\pi, \pi/2, 1/2)$ for simulation V44 at $t=12$. The volume rendering displays the streamwise velocity $u_x$, and also shown are field lines for the total field $\bm B = \bm B_0 + \bm b$ colored using $|\bm B|$ to denote the local magnetic field intensity. Note that there is a flux of magnetic field across the boundary.}
	\label{results:fig:vacuum}
\end{figure}

\begin{table*}
	\begin{center}
		\small
		\begin{tabular}{c c c c c S[table-format=-1.2e-1] S[table-format=-1.1e-1] S[table-format=-1.1e-1]}
			\toprule
			\toprule
			Run ID	& $N_x\times N_y\times N_z$ & ${Ha}$ & ${Re_0}$  & ${R}$ &  $b_{0}$ &  ${G}$   &${\nu=\eta}$\\
			\midrule
V1  & $128 \times  64 \times  231$ &    1 &  116 &  116 &    1e-2 &    5e-2 &    5e-3\\
V2  & $128 \times  64 \times  231$ &  2.5 &   85 &   34 &  2.5e-2 &    5e-2 &    5e-3\\
V3  & $128 \times  64 \times  231$ & 3.75 &   64 &   17 & 3.75e-2 &    5e-2 &    5e-3\\
V4  & $128 \times  64 \times  231$ &    5 &   49 &   10 &    5e-2 &    5e-2 &    5e-3\\
V5  & $128 \times  64 \times  231$ &  7.5 &   33 &    4 &  7.5e-2 &    5e-2 &    5e-3\\
V21 & $128 \times  64 \times  231$ &  2.5 &  170 &   68 & 1.25e-2 &  2.5e-2 &  2.5e-3\\
V22 & $128 \times  64 \times  231$ &  2.5 &  405 &  162 &    5e-3 &    1e-2 &    1e-3\\
V23 & $128 \times  64 \times  231$ &  2.5 & 1648 &  659 & 1.25e-3 &  2.5e-3 &  2.5e-4\\
V24 & $256 \times 128 \times  487$ &  2.5 & 3198 & 1279 &    5e-4 &    1e-3 &    1e-4\\
V41 & $128 \times  64 \times  231$ &    5 &   91 &   18 &  2.5e-2 &  2.5e-2 &  2.5e-3\\
V42 & $128 \times  64 \times  231$ &    5 &  215 &   43 &    1e-2 &    1e-2 &    1e-3\\
V43 & $128 \times  64 \times  231$ &    5 &  824 &  165 &  2.5e-3 &  2.5e-3 &  2.5e-4\\
V44 & $256 \times 128 \times  487$ &    5 & 2170 &  434 &    1e-3 &    1e-3 &    1e-4\\
            \bottomrule
            \bottomrule
        \end{tabular}
    \caption{Summary of the simulations performed for the vacuum boundary conditions case. All the runs employ a number of continuation points $C_z = 25$, as well as 9\textsuperscript{th} order FC-Gram extensions ($d_z = 10$). In all the cases the fluid spawns the domain $L_x \times L_y \times L_z = 2\pi \times \pi \times 1$. As in Table \ref{results:tbl:conducting}, ``Run ID" labels each simulation, $N_x\times N_y\times N_z$ gives the linear resolution in all directions, $Ha$ is the Hartman number, $Re_0$ the Reynolds number, $R$ the modified Reynolds number, $b_0$ the amplitude of the external uniform magnetic field, $G$ the external pressure gradient, and $\nu=\eta$ are the dimensionless kinematic viscosity and magnetic diffusivity.}
    \label{results:tbl:vacuum}
    \end{center}
\end{table*}
We now consider the simulations performed for the case of vacuum surroundings. A summary of all the simulations performed for this case can be found in \cref{results:tbl:vacuum}. Also, a graphical representation of the 3D fields, obtained using the software \texttt{VAPOR} \cite{Clyne2007,Vapor}, is shown in \cref{results:fig:vacuum}. Note that in this case field lines correspond to the total magnetic field $\bm B = b_0 \bm{\hat z} + \bm b$ and that the induced magnetic field $\bm b$ is non-zero outside the computational domain.

\begin{figure}[t]
	\centering
	\includegraphics[width=.8\linewidth, keepaspectratio]{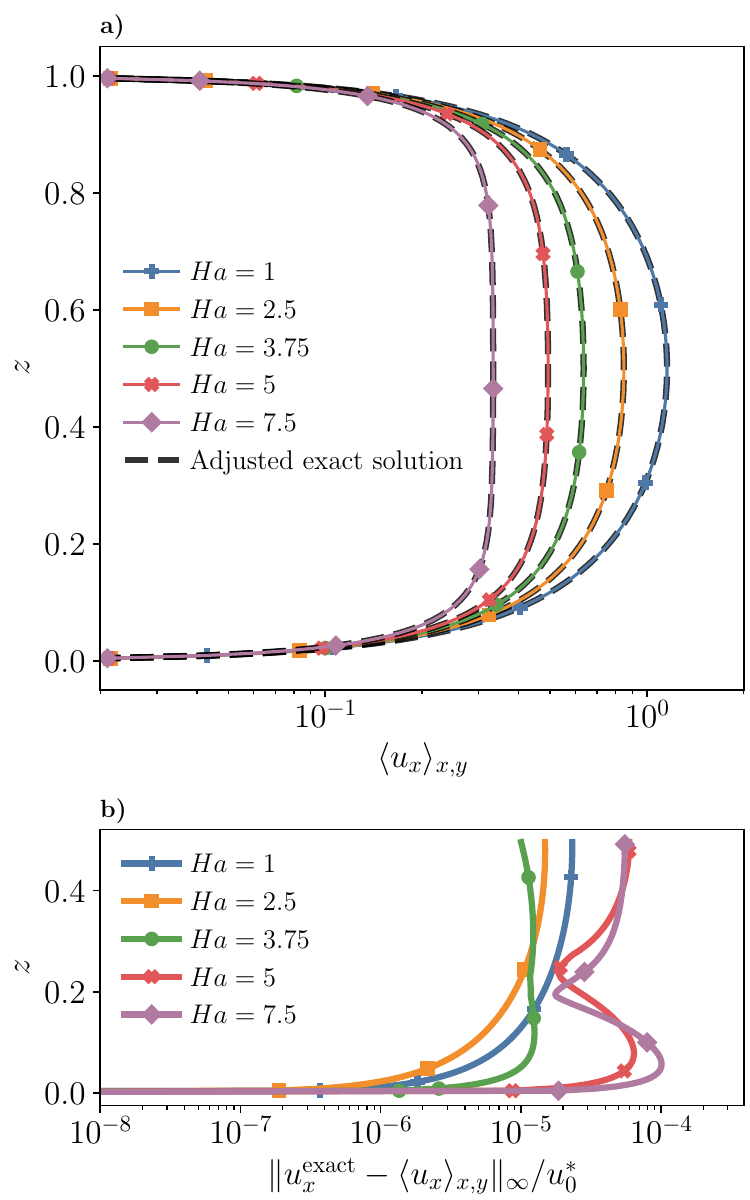}
	\caption{Results for simulations V1 to V5 after 200 eddy turnover times. \textbf{a)} Mean vertical profile for the streamwise velocity $\langle u_x \rangle_{x,y}$, alongside black dashed lines which represent the corresponding closed-form laminar solution for each value of Ha, with $u_0$ adjusted so that the profile's centerline value is matched. \textbf{b)} Relative maximum difference between the mean profiles and the laminar solution as a function of $z$. The scaled value $u_0^*$ is used as normalization. For both panels, profiles for varying values of $Ha$ are denoted employing different colors and markers.}
	\label{results:fig:vacuum_varying_b0}
\end{figure}

As before, we start by studying laminar scenarios, in which $Ha$ is of order 1, and modest values of the Reynolds number are considered. In this region of parameter space, the effect of varying $Ha$ from 1 to 7.5 is explored by changing the magnitude of the externally imposed magnetic field, resulting in runs V1 to V5. Also as before, we start the system from an initial state of random velocity and vector potential fields whose energy is $10^{-2}$, and is concentrated at the largest scales of the system. In \cref{results:fig:vacuum_varying_b0} we show the mean streamwise velocity profile after 200 eddy turnover times for this set of simulations, and compare it to the analytical solution given by \cref{results:eq:u}. Note, however, that contrary to the conducting case a theoretical expression for $u_0$ is not available, and hence the laminar profile is normalized so that its maximum matches the one obtained from the simulations. Similarly to the conducting case, we find a very good agreement between the normalized analytical solution and the obtained profiles. This is also quantified in \cref{results:fig:vacuum_varying_b0}, where panel \textbf{b)} shows the average absolute difference between analytical and simulated profiles, obtaining relative errors in the range $10^{-6}$--$10^{-4}$, comparable to those obtained for the conducting case. For all the simulations, standard deviations in the $x$ and $y$ directions are of order $10^{-8}$, indicating the homogeneity of the flow in the streamwise and spanwise directions.

We now focus our attention on simulations V2X and V4X, where we gradually increase the values of $Re_0$ and $R$ at fixed $Ha$. As before, we consider the cases $Ha=2.5$ and $Ha=5$, as they represent qualitatively different Hartmann boundary layer thickness for the chosen geometry. To explore distinct values of $R$, starting from runs V2 and V4, we add 10\% of random noise to the last output of the available simulation with the highest $R$ value while decreasing the viscosity, the forcing, and the applied magnetic field. As it was the case for simulations with conducting walls, each time the new parameters obey $\nu' = \nu/2$, $G' = G/2$ and $b_0' = b_0/2$, where primed variables denote quantities for the larger $R$ simulation. This allows for easy comparison between simulations with conducting walls and with vacuum surroundings.

\begin{figure}[t]
	\centering
	\includegraphics[width=.8\linewidth, keepaspectratio]{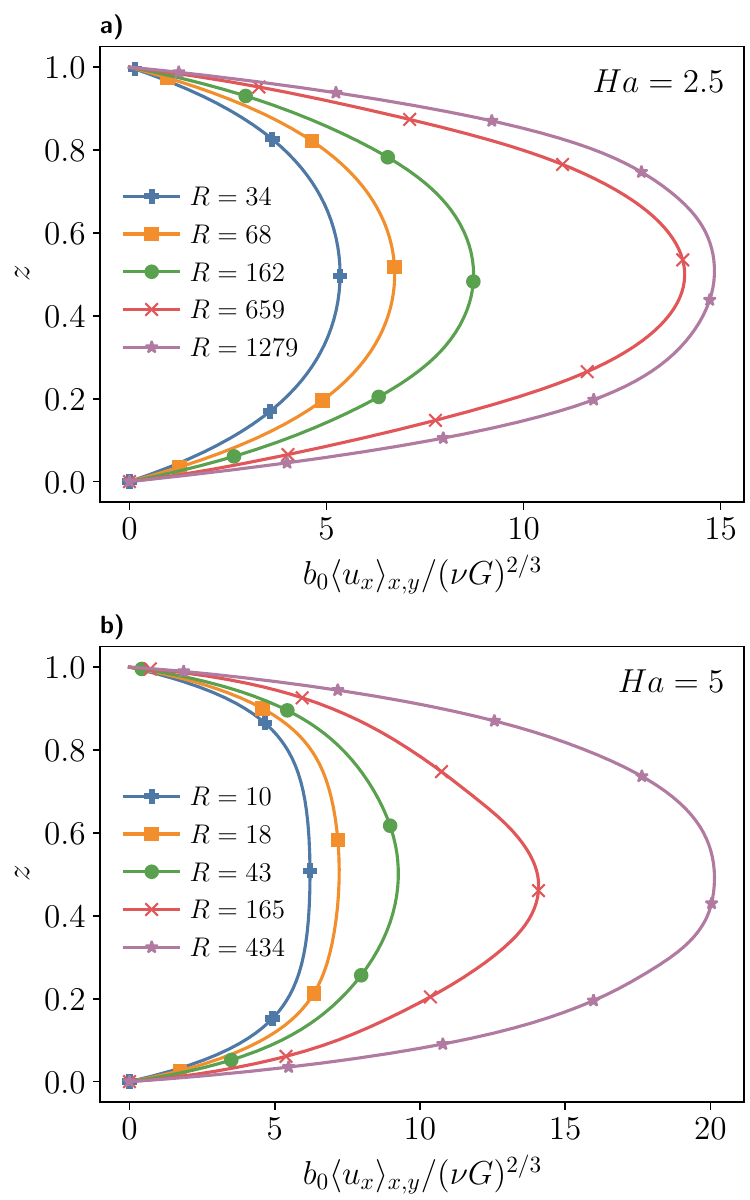}
	\caption{\textbf{a)} Mean vertical profile of the streamwise velocity normalized by a prescription for $u_0$, $u_0 \propto (\nu G)^{2/3}/b_0$, as a function of $z$ for the set of simulations V2X ($Ha=2.5$). \textbf{b)} Same for simulations V4X ($Ha=5$). In all the cases, profiles for different values of $R$ are differentiated by their colors and markers.}
	\label{results:fig:vacuum_fixed_ha}
\end{figure}

In \cref{results:fig:vacuum_fixed_ha} we show the profiles obtained for the mean streamwise velocity, now normalized by $b_0/(\nu G)^{2/3}$. This normalization follows from dimensional analysis and from the observation that $u_0 \propto b_0^{-1}$ in \cite{Chang1961,Vantieghem2009} for insulating boundaries at $Pm \ll 1$, a situation somewhat closer to the vacuum surroundings we consider here than the theoretical predictions for a perfect conductor. \Cref{results:fig:vacuum_fixed_ha} shows normalized velocity profiles which display larger centerline values as $R$ is increased. Interestingly, not only the intensity of the streamwise velocity is modified, but also its profile, with more parabolic profiles as $R$ increases. This could suggest that in the $Ha \approx 1$ and $Pm = 1$ regime, a more Poiseuille-like solution is recovered for $Re_0 > 1$ in the case of vacuum surroundings, although further research is required to confirm this observation.

\begin{figure}[t]
	\centering
	\includegraphics[width=\linewidth, keepaspectratio]{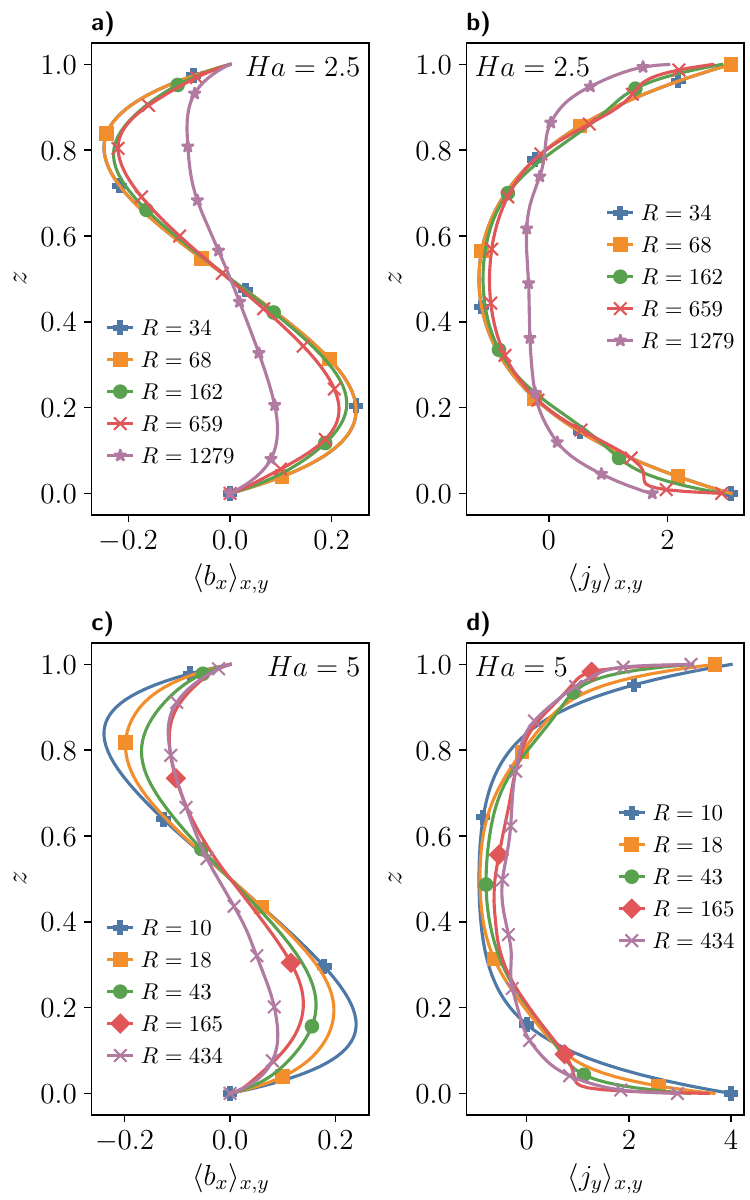}
	\caption{Left column: mean vertical profile for the streamwise magnetic field $\langle b_x \rangle_{x,y}$ as a function of $z$ for simulations V2X (a) and V4X (c). Right column: mean vertical profile for the spanwise current density field $\langle j_y \rangle_{x,y}$ as a function of $z$ for simulations V2X (b) and V4X (d). In all the cases profiles for different values of $R$ are differentiated by their colors and markers.}
	\label{results:fig:vacuum_magnetic}
\end{figure}

Finally, \Cref{results:fig:vacuum_magnetic} displays the vertical mean profiles obtained for the streamwise magnetic field and the spanwise current. Note the differences in the values of the fields in the boundaries when compared with the perfectly conductor. In panels \textbf{a)} and \textbf{c)} it can be seen that, as it was for the conducting case, the maximum value of the streamwise magnetic field diminishes as $Ha$ increases, although contrary to the findings of \cref{sec:results:conducting}, that maximum value is not attained at the walls. For the spanwise current density, shown in panels \textbf{b)} and \textbf{d)}, the vertical profile showcases a similar shape to the one found for the conducting case. However, as boundary conditions do not require $\bm j \times \hat{\bm z} = \bm 0$, a significant positive current is found near the walls, which diminishes towards the core and becomes negative sufficiently far away from the Hartmann layer. Minimum values for the current density are similar for both sets of boundary conditions.

How well the boundary conditions are satisfied, and how these numerical solutions converge as spatial resolution is increased or the time step is decreased, is considered in detail in the next section.

\section{Error estimation and convergence}
\label{sec:convergence}

\begin{figure}
	\centering
	\includegraphics[width=\linewidth, keepaspectratio]{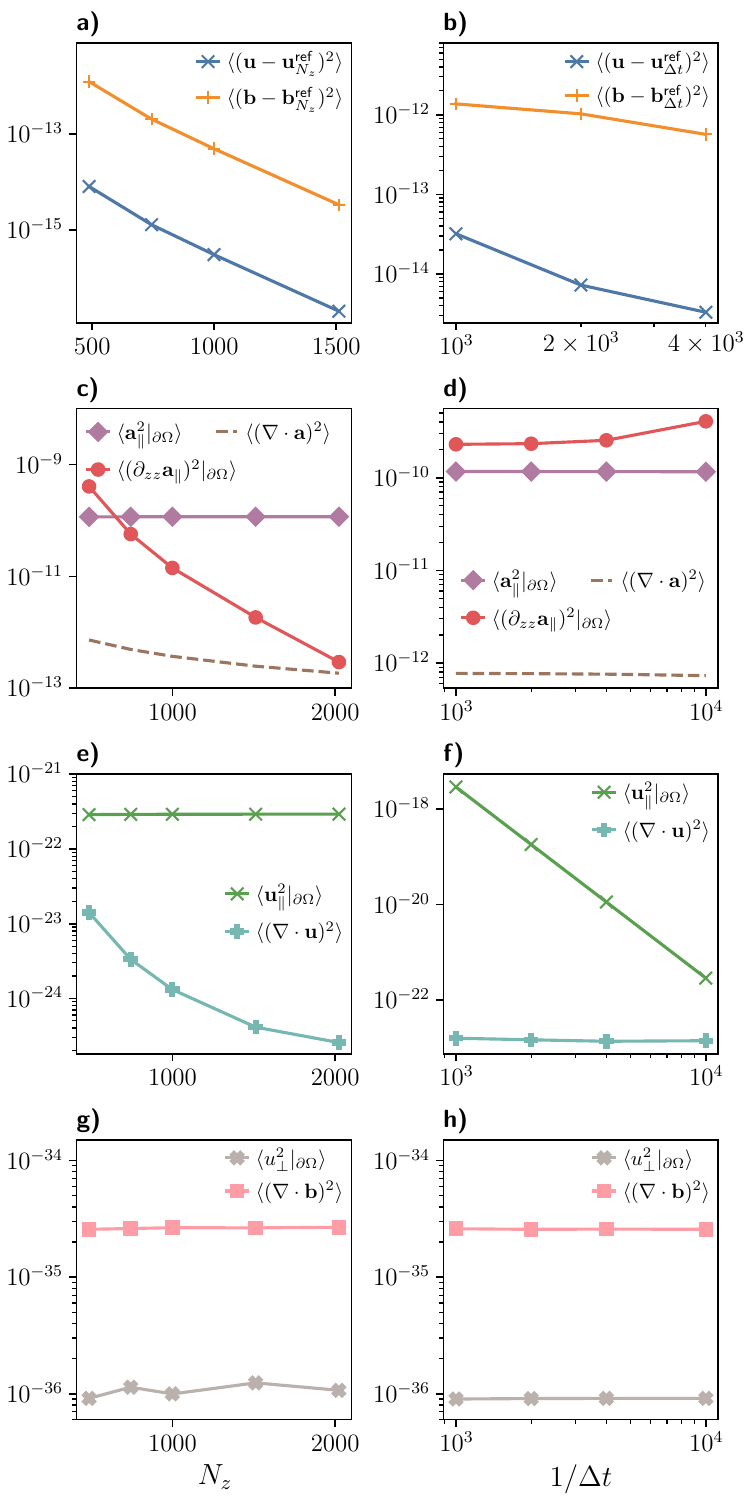}
	\caption{Errors for the perfectly conducting solver for varying the spatial vertical (first column, in semilog scale) and temporal (second column, in log-log scale) resolutions. In \textbf{a)} and \textbf{b)} the bulk error for $\bm{u}$ and $\bm{b}$ is shown, considering the solution at the highest (space or time) resolution as reference. Panels \textbf{c)} and \textbf{d)} display, as a function of the spatial or temporal resolution, the mean square error of both the boundary conditions for $\bm{a}$ and its gauge imposition. Similarly, in \textbf{e)} and \textbf{f)} the error in $\bm \nabla \cdot \bm u$ is exhibited, together with error in the tangential velocity at the wall. In the bottom row, \textbf{g)} and \textbf{h)} show the accuracy of $\bm \nabla \cdot \bm b$ and the normal velocity at the wall. All quantities reported are mean squared errors. Simulations with varying vertical resolution where carried out with $\Delta t=1\times10^{-4}$, whereas $N_z=487$ was employed for runs with changing time steps.}
	\label{convergence:fig:convergence}
\end{figure}

For the two Hartmann setups presented in the previous section we now study the numerical performance of the method. As we use a standard Fourier pseudo-spectral method in the periodic $\hat{\bm x}$ and $\hat{\bm y}$ directions, the results follow the well known exponential convergence when spatial resolution ($N_x$ and $N_y$) is varied in those directions \cite{Canuto2006}. Hence, we focus instead on varying the resolution in the $\hat{\bm z}$ direction ($N_z$), and on the effect of changing the time step ($\Delta t$). Let's summarize what we will show in this section: First, we remind the reader that we have physical fields, and auxiliary fields that depend on the gauge freedom. Second, we will consider two types of errors: errors in the bulk solution of the physical fields as $N_z$ or $\Delta t$ is changed, and errors in the boundary conditions also as the two parameters are changed. Both kinds of errors will be considered using mean squared values. Third, we will see there are three types of behaviors that follow directly from the implementation of the method: some errors scale with $N_z$, some with $\Delta t$, and some conditions are either auxiliary or so well satisfied that they are very small and constant, in many cases of the order of the machine truncation error.

In particular, for both types of boundary conditions, the convergence of the solution as the resolution increases is analyzed by considering a simulation at large resolution as a reference field, and estimating mean squared errors for lower resolution solutions. The resulting error decreases very fast, especially as a function of the spatial resolution. We look into the accuracy to which the method preserves the physical restrictions $\bm \nabla \cdot \bm u = \bm \nabla \cdot \bm b = 0$, which are seen to be tiny in all the cases, particularly for $\bm \nabla \cdot \bm b$ for which the error is kept $\mathcal{O}(10^{-28})$ or smaller in all cases. This is important as small numerical violations of this condition were shown to have a measurable impact in the physical solutions of the MHD equations \cite{Brackbill1980,Balsara2004}. The gauge condition $\bm \nabla \cdot \bm a = 0$ is also seen to remain overall small. 

\begin{figure}
	\centering
	\includegraphics[width=\linewidth, keepaspectratio]{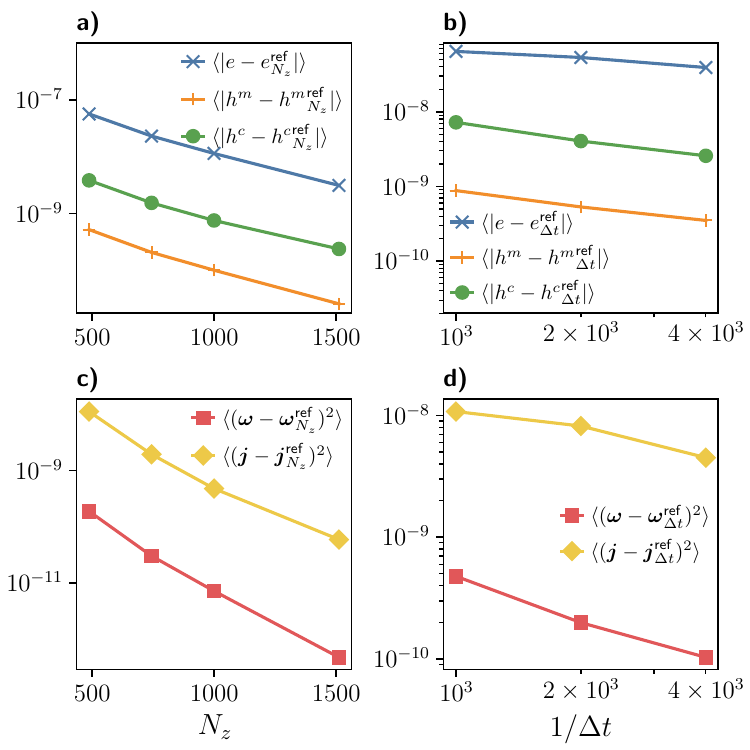}
	\caption{\textbf{a)} and \textbf{b)}: mean absolute error for the total energy density $e$ (in blue, with $\mathsf x$), the magnetic helicity density $h^m$ (in orange, with $\mathsf +$), and the cross helicity density $h^c$ (in green, with $\circ$) in the bulk of the fluid, as a function of (\textbf{a}) the vertical spatial resolution $N_z$, and (\textbf{b}) the inverse of the temporal resolution $1/\Delta t$. \textbf{c)} and \textbf{d)}: mean quadratic error for the vorticity $\bm \omega$ (in red, with $\square$) and current density $\bm j$ (in yellow, with $\Diamond$), also in the bulk of the fluid, and as a function of (\textbf{c}) the spatial and (\textbf{d}) the inverse of the temporal resolution. For all quantities the reference simulations (``ref'' superscript) are the same as those in \cref{convergence:fig:convergence}.}
	\label{convergence:fig:pointwise}
\end{figure}

Finally, we consider the accuracy in the boundary conditions. Regarding the boundary conditions for the velocity field, we obtain results comparable to \cite{Fontana2020}, as both the tangential and normal velocity remain very small and controlled by the time integration. Concerning electromagnetic boundary conditions, in the perfect conductor the error in $(\partial^2_{zz} \bm{a}_{\parallel})|_{\partial \Omega} = 0$ decreases fast with increasing $N_z$, and the error in $(\bm{a}_{\parallel})|_{\partial \Omega} = 0$ remains small and $\mathcal{O}(10^{-10})$. All other conditions associated to gauge freedoms, even though they don't affect the physical solution, remain $\mathcal{O}(10^{-12})$ or smaller. Finally, for vacuum surroundings, the error in the physically relevant boundary condition $(\partial_{z}\bm a_{\parallel} + \gamma_{mn}\bm a_{\parallel})|_{\partial \Omega} = 0$ decreases with $N_z$, while the error in the gauge condition $(\partial_{z} a_\perp + \gamma a_\perp)|_{\partial \Omega}=0$ is controlled by the temporal resolution. All these dependencies of errors with $N_z$ or $\Delta t$ are to be expected from the implementation of the conditions in the split time stepping, depending on whether they are applied at either the intermediate step, the Poisson solver and spectral projection, or the final step.

The details of the errors for the perfectly conducting solver are reported in \cref{convergence:fig:convergence}. In the figure, panels \textbf{a)} and \textbf{b)} show the mean pointwise square error when comparing $\bm u$ and $\bm b$ to those obtained from the simulation at the highest resolution, that is $\langle(\bm u - \bm u^\text{ref})^2\rangle_{x,y,z}$ (with the analogous expression for $\bm b$). When analyzing spatial convergence the reference fields (denoted with $^\text{ref}_{N_z}$) employ $N_z=2023$ vertical grid points, whereas the reference to study the effect of varying the time resolution (denoted $^\text{ref}_{\Delta t}$) utilizes $\Delta t=10^{-4}$. A clear convergence for increasing resolution is observed in both cases. In particular, note that when varying $N_z$ by only a factor of $\approx 3$, a decrease in the error by almost 3 orders of magnitude is observed, compatible with the fast convergence reported for a purely hydrodynamic FC-Gram pseudo-spectral method in \cite{Fontana2020}.

\begin{figure}
	\centering
	\includegraphics[width=\linewidth, keepaspectratio]{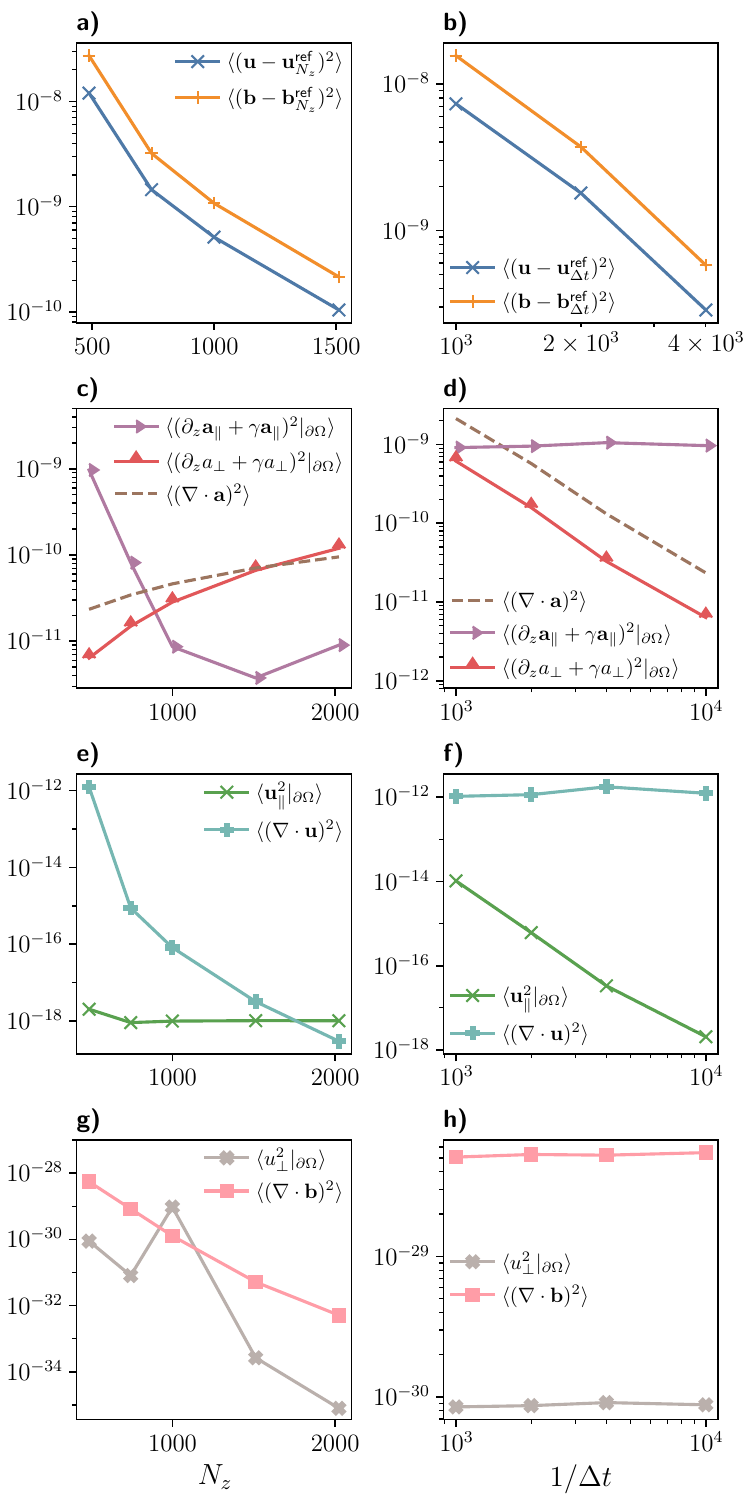}
	\caption{Errors for the vacuum solver for varying spatial vertical (first column, in semilog scale) and temporal (second column, in log-log scale) resolutions. In \textbf{a)} and \textbf{b)} the bulk error for $\bm{u}$ and $\bm{b}$ solutions is shown, considering the solution at the highest (spatial or temporal) resolution as reference. Panels \textbf{c)} and \textbf{d)} display, as a function of the spatial and temporal resolution, the average error of both the boundary conditions for $\bm{a}$ and its gauge imposition. Similarly, in \textbf{e)} and \textbf{f)} the error in $\bm \nabla \cdot \bm u$ is exhibited, together with the error in the tangential velocity at the wall. In the bottom row, \textbf{g)} and \textbf{h)} show the accuracy of $\bm \nabla \cdot \bm b$ and the normal velocity at the wall. All quantities reported are mean squared errors. Simulations with varying vertical resolution where carried out with $\Delta t=1\times10^{-4}$, whereas $N_z=487$ was employed for runs with changing time resolution.}
	\label{convergence:fig:convergencei}
\end{figure}

In \cref{convergence:fig:convergence} \textbf{c)} and \textbf{d)}, the mean square values of both $\bm a_\parallel$ and $\partial^2_{zz} \bm a_\parallel$ at the wall are shown. As expected, $\langle (\partial^2_{zz} \bm a_\parallel)^2\rangle$ is seen to rapidly decrease with increasing $N_z$ (note it is FC-Gram imposed at the end of each time step), while it remains approximately constant with $\Delta t$. Meanwhile, $\langle \bm a_\parallel^2 \rangle$ remains small albeit independent of $N_z$ and $\Delta t$ (the condition is enforced in the middle of the time step, over $\bm a^*_\parallel$, and affected by the later gauge imposition). Also in \textbf{c)} and \textbf{d)} we show the mean square value of $\bm \nabla \cdot \bm a$, which decreases as a function of $N_z$ down to $\mathcal{O}(10^{-13})$, and remains constant when varying $\Delta t$. The remaining magnetic boundary condition, $(\partial_{z} a_\perp)|_{\partial \Omega}=0$, which is only imposed to favor the stability of the method as well as to help enforce $\bm \nabla \cdot \bm a = 0$ at the wall, remains accurately enforced, with typical mean quadratic values of $10^{-14}$.

Panels \textbf{e)} and \textbf{f)} in \cref{convergence:fig:convergence} display the average quadratic errors $\langle (\bm \nabla \cdot \bm u)^2\rangle_{x,y,z}$, and $\langle \bm u_\parallel^2 \rangle_{x,y}$ at the wall. The first one decreases about two orders of magnitude when increasing $N_z$ from 487 to 2023 (and independent of $\Delta t$), while the latter remains essentially constant and $\mathcal{O}(10^{-21})$ as $N_z$ is increased, and decreases as $(1/\Delta t)^{-4}$ when the time resolution is varied. Finally, panels \textbf{g)} and \textbf{h)} show the mean square error in the wall-normal velocity and in $\bm \nabla \cdot \bm b$ everywhere. As both are strictly imposed (the former by the Neumann boundary conditions in the Poisson equation for the pressure, and the latter by construction as $\bm b = \bm \nabla \times \bm a$), errors are $\mathcal{O}(10^{-34})$ or smaller. It can be readily observed that both conditions are satisfied to the accuracy allowed by double precision arithmetic. As previously mentioned, this is one of the main advantages of our method, as mass conservation and keeping the magnetic field solenoidal are relevant for properly capturing the physics of the problem \cite{Balsara2004}.

Also important is the numerical accuracy in the spatial densities of the total energy $e = \bm u^2 + \bm b^2$, the magnetic helicity $h^m = \bm a \cdot \bm b$, and the cross helicity $h^c = \bm u \cdot \bm b$. As previously mentioned, an accurate determination of these quantities is very important in the study of MHD, and for the flow evolution. In particular, the integral magnetic helicity displays an inverse cascade and is consequently important for the evolution of the large scales of the flow. On the contrary, the total energy and the cross-helicity are forward-cascading magnitudes in MHD, and thus transfer energy to the small scale structures such as current sheets and vortex filaments, as well as possibly generating Alfvénic states. In panels \textbf{a)} and \textbf{b)} of \cref{convergence:fig:pointwise}, the mean of the pointwise absolute errors in $e$, $h^c$, and $h^m$ are shown compared to a high resolution simulation, both as a function of $N_z$ and $\Delta t$. In each case the reference simulation is the same as that used for \cref{convergence:fig:convergence}. Convergence is observed when both $N_z$ is increased or $\Delta t$ is decreased, most notably in the former case where a reduction of more than an order of magnitude for all the aforementioned quantities is observed when increasing the spatial resolution threefold. This is to be expected as pointwise errors in these quantities are governed by the spatial discretization.

A more direct measure of the small scales can be obtained by studying the vorticity $\bf \omega$ and current density $\bf j$, which also control the volumetric dissipation rate of the ideal invariants. The mean of the pointwise quadratic error as a function of $N_z$ and $\Delta t$ under the proposed solver for the conducting case is displayed in panels \textbf{c)} and \textbf{d)} of \cref{convergence:fig:pointwise}, by comparing again against the high resolution reference simulation. In both cases good convergence properties are observed, particularly in the spatial case, where the error is reduced by more than two orders of magnitude when increasing $N_z$ by a factor of three.

The same analysis as the one shown in \cref{convergence:fig:convergence} for the perfectly conducting solver is shown in \cref{convergence:fig:convergencei} but now for the vacuum solver. As before, panels \textbf{a)} and \textbf{b)} display the mean pointwise square error in $\bm u$ and $\bm b$ everywhere in the domain for increasing resolutions. The reference solutions employ $N_z=2023$ vertical grid points for examining the dependence on the spatial resolution, and $\Delta t=10^{-4}$ for studying the effect of the timestep size. In both cases a steep convergence is found for increasing resolution. Also, a $(1/\Delta t)^{-4}$ trend is observed when varying $\Delta t$, compatible for the quadratic errors of a 2nd order Runge-Kutta method. For the vertical spacing, a decrease of the errors by two orders of magnitude is found when spatial resolution is changed by a factor of $\approx 3$. As before, for the solenoidal and boundary conditions we see in the rest of the panels the dependence of errors on either $N_z$ or $\Delta t$ characteristic of time splitting methods. More precisely, in panels \textbf{c)} and \textbf{d)}, the mean squared error of $\langle (\partial_z \bm a_\parallel + \gamma \bm a_\parallel)^2\rangle_{x,y}$ at the wall is observed to decrease by two orders of magnitude when increasing $N_z$, whereas both $\langle (\partial_z \bm a_\perp + \gamma a_\perp)^2 \rangle_{x,y}$ at the wall and $\langle (\bm \nabla \cdot \bm a)^2\rangle_{x,y,z}$ everywhere decrease with decreasing $\Delta t$, with a uniform $(1/\Delta t)^{-2}$ convergence rate.

Also in \cref{convergence:fig:convergencei}, panels \textbf{e)} and \textbf{f)} exhibit the same behavior than for the conducting solver, with the slip velocity at the wall being regulated by the timestep size, and the error in the divergence of the velocity field depending on $N_z$. A detail worth pointing out is the contrasting behavior between $(\bm \nabla \cdot \bm u)^2$ and $(\bm \nabla \cdot \bm a)^2$. The first one is lower overall and regulated by the spatial resolution, whereas the latter is dependent on the temporal discretization. This difference is by design, as $\bm \nabla \cdot \bm u$ is of greater physical significance, being the divergence of a physical field, while the condition on $\bm a$ is a choice of gauge. Finally and more importantly, as it was also the case for the conducting solver, panels \textbf{g)} and \textbf{h)} show that the errors in the wall normal velocity and in $\bm \nabla \cdot \bm b$ everywhere are very small, of $\mathcal{O}(10^{-28})$ or smaller. Finally, the pointwise behavior of $e$, $h^m$, $h^c$, $\bf \omega$, and $\bf j$ was also analyzed (not shown), and results were found to be similar to those found for the conducting case (\cref{convergence:fig:pointwise}) as the temporal and spatial resolutions were increased.

\section{Discussion}
\label{sec:conclusions}

A novel parallel pseudo-spectral method was presented, based on FC-Gram transforms to evolve the incompressible MHD equations in cuboid non-periodic domains. The method can be easily generalized to other (e.g., compressible) MHD systems, as well as to other geometries, as e.g., the case of the radial coordinate in a sphere or a spherical shell as often considered in geodynamo simulations. As a result, it can be relevant for multiple problems found in geophysics, industrial cooling systems, and space physics.

Being Fourier-based, and due to the resulting dispersionless computation of spatial derivatives, the method is high-order and it has no spurious dispersion (nor ``pollution") errors that usually arise in finite differences or finite element methods. Additionally, the Fourier base allows for fast and efficient solution of Poisson equations, as those arising when imposing solenoidal conditions for the fields, or when solving for the electric scalar potential in other formulations of plasma flows. Also, as a result of the properties of the FFTs, is computationally efficient (with $\mathcal{O}(N\log N)$ operations per Fourier transform) and can be effectively parallelized to run in computer clusters. Overall, the method inherits many of the properties of classical pseudo-spectral solvers in periodic domains, with a small computational overhead to compute continuations at the boundaries and impose boundary conditions. The method is also well conditioned: differential operators are diagonal and easy to invert (compared with other spectral methods using, e.g., Chebyshev bases \cite{Julien2009}), and all ill-conditioning reduces to the computation of small tables which can be computed beforehand with arbitrary precision, and stored.

An important feature of the method is that it evolves the vector potential (allowing direct computation of physically relevant quantities such as the magnetic helicity), and that it keeps the solenoidal condition on the magnetic field with errors of $\mathcal{O}(10^{-28})$ or smaller, of physical relevance as small errors in this condition in low order numerical methods have been shown to have a measurable impact in physical solutions \cite{Brackbill1980,Balsara2004}.

The method can be easily extended to consider two-fluid descriptions of plasmas, as, e.g., the Hall-MHD equations. These equations can be written using the vector potential and the Coulomb gauge (which in this formulation also improves the numerical stability \cite{Mininni2005b}), and the high-order convergence in the estimation of spatial derivatives becomes useful for the computation of high-order derivatives of nonlinear terms in these equations. For a discussion of Hall-MHD equations in the vector potential formulation in pseudospectral methods (for the case of periodic boundary conditions), see \cite{Mininni2005b}.

Two boundary conditions relevant in geophysics and astrophysics were explicitly considered: perfectly conducting walls, and vacuum surroundings. These conditions arise, e.g., in geodynamo simulations when considering respectively the interface of the liquid core with the inner solid core, and the interface of the liquid core with the outter mantle \cite{Jones2000,Roberts2000}. Boundary conditions for the vector and electrostatic potentials where derived for these scenarios, resulting in Robin and second normal derivative conditions on the fields. The case of Robin boundary conditions is of practical interest as it also arises in many other electromagnetic problems. As a result, we derived a FC-Gram method to deal with these boundary conditions with high precision and minimal overhead.

An explicit time-splitting technique was presented for the temporal integration, allowing for imposition of either solenoidality, intermediate, or physical boundary conditions at every substep. Explicit time stepping strategies as the ones presented here are particularly suitable for high Reynolds regimes \cite{Kim1985,Orszag1986}, as the time step is dominated by the CFL condition on the advection term, specially in uniform grids. For cases with very thin boundary layers, locally refined meshes could be easily employed, as the FC-Gram method allows matching continuously spectral approximations for different spatial regions.

The full numerical method was implemented in an existing fully parallelized and fluid-oriented PDE solver, \texttt{SPECTER}, and made openly available. This implementation was used to validate the method and study its convergence and numerical errors considering a paradigmatic wall-bounded MHD problem, namely a Hartmann flow. Both types of boundary conditions were tested, obtaining results compatible with previous studies. Errors were shown to remain small in all cases, and to decrease either with increasing spatial resolution or time resolution depending on the implementation of the different conditions in the time-splitting method. For the future, we plan to extend \texttt{SPECTER} to provide other electromagnetic solvers of interest for geophysics and astrophysics using the FC-Gram pseudo-spectral method, including MHD flows with convection, or compressible MHD solvers.

\section*{Declaration of competing interest}
The authors declare that they have no known competing financial interests or personal relationships that could have appeared to influence the work reported in this paper.

\section*{Acknowledgments}
The authors acknowledge support from CONICET and ANPCyT through PIP, Argentina Grant No.~11220150100324CO, and PICT, Argentina Grant No.~2018-4298. This work was also supported by National Science Foundation, USA through contract DMS-1714169, and by the NSSEFF Vannevar Bush Fellowship, USA under contract number N0001416-1-2808. We also thank the Physics Department at the University of Buenos Aires for providing computing time on its Dirac cluster.
\vskip 1cm

\bibliography{specter_mhd}
\bibliographystyle{elsarticle-num}

\end{document}